\documentclass{article} 
\usepackage{iclr2023_conference,times}

\usepackage{amsmath, bm, amssymb, amsthm}
\usepackage{multirow}
\usepackage{mathtools}
\usepackage{threeparttable}
\usepackage{tablefootnote}
\usepackage{xspace}
\usepackage{graphicx}
\usepackage{subfigure}

\usepackage[utf8]{inputenc} 
\usepackage[T1]{fontenc}    
\definecolor{mypink}{RGB}{253,111,150}
\definecolor{mydarkblue}{rgb}{0,0.08,0.45}

\usepackage[colorlinks,
    linkcolor=mydarkblue,
    anchorcolor=mydarkblue,
    urlcolor=mypink,
    citecolor=mydarkblue,]{hyperref} 
\usepackage{hyperref}       
\usepackage{cleveref}
\usepackage{url}            
\usepackage{booktabs}       
\usepackage{amsfonts}       
\usepackage{nicefrac}       
\usepackage{microtype}      
\usepackage{algorithm}
\usepackage{algorithmic}
\usepackage{wrapfig,tikz}
\usepackage{chemformula}
\usepackage{adjustbox}
\usepackage[normalem]{ulem}
\usepackage[textwidth=0.1in,textsize=footnotesize]{todonotes}

\usepackage{amsmath,amsfonts,bm}

\def\eqref#1{equation~\ref{#1}}

\def\1{\bm{1}}

\def\va{{\bm{a}}}
\def\vb{{\bm{b}}}

\def\vk{{\bm{k}}}

\def\vm{{\bm{m}}}

\def\vq{{\bm{q}}}
\def\vr{{\bm{r}}}
\def\vs{{\bm{s}}}

\def\vv{{\bm{v}}}

\def\vx{{\bm{x}}}

\def\vz{{\bm{z}}}

\def\mI{{\bm{I}}}

\def\mO{{\bm{O}}}

\def\mR{{\bm{R}}}

\DeclareMathAlphabet{\mathsfit}{\encodingdefault}{\sfdefault}{m}{sl}
\SetMathAlphabet{\mathsfit}{bold}{\encodingdefault}{\sfdefault}{bx}{n}

\def\gF{{\mathcal{F}}}

\def\gL{{\mathcal{L}}}

\def\gO{{\mathcal{O}}}
\def\gP{{\mathcal{P}}}

\def\gS{{\mathcal{S}}}
\def\gT{{\mathcal{T}}}

\def\gX{{\mathcal{X}}}

\def\sR{{\mathbb{R}}}

\title{Protein Sequence and Structure Co-Design \\with Equivariant Translation}

\newcommand{\method}{\textsc{ProtSeed}\xspace}
\newcommand{\rosetta}{\textsc{RAbD}\xspace}
\newcommand{\gnn}{\textsc{GNN}\xspace}
\newcommand{\diffusion}{\textsc{Diffusion}\xspace}
\newcommand{\afdesign}{\textsc{AFDesign}\xspace}
\newcommand{\gvp}{\textsc{GVP-GNN}\xspace}
\newcommand{\esmif}{\textsc{GVP-Transformer}\xspace}
\newcommand{\structgnn}{\textsc{Structured GNN}\xspace}

\newtheorem{proposition}{Proposition}

\author{
Chence Shi$^{1,2,3}$, Chuanrui Wang$^{2,3}$, Jiarui Lu$^{2,3}$, Bozitao Zhong$^{2,3}$, Jian Tang$^{1,2,4,5}$\\
$^1$BioGeometry
$^2$Mila - Qu\'ebec AI Institute 
$^3$Universit\'e de Montr\'eal \\
$^4$HEC Montr\'eal 
$^5$CIFAR AI Research Chair \\
\texttt{chence.shi@\{biogeom.com,umontreal.ca\}}\\
\texttt{\{chuanrui.wang,jiarui.lu,bozitao.zhong\}@umontreal.ca} \\
\texttt{jian.tang@hec.ca}
}

\newcommand{\wcr}[1]{{\color[rgb]{0.8, 0.1, 0.1}{WCR: #1}}}

\iclrfinalcopy 
\begin{document}

\maketitle

\begin{abstract}

Proteins are macromolecules that perform essential functions in all living organisms.
Designing novel proteins with specific structures and desired functions has been a long-standing challenge in the field of bioengineering.
Existing approaches generate both protein sequence and structure using either autoregressive models or diffusion models, both of which suffer from high inference costs. 
In this paper, we propose a new approach capable of protein sequence and structure co-design, which iteratively translates both protein sequence and structure into the desired state from random initialization, based on context features given \textit{a priori}.
Our model consists of a trigonometry-aware encoder that reasons geometrical constraints and interactions from context features, and a roto-translation equivariant decoder that translates protein sequence and structure interdependently. 
Notably, all protein amino acids are updated in one shot in each translation step, which significantly accelerates the inference process.
Experimental results across multiple tasks show that our model outperforms previous state-of-the-art baselines by a large margin, and is able to design proteins of high fidelity as regards both sequence and structure,
with running time orders of magnitude less than sampling-based methods.

\end{abstract}
\section{Introduction}
\label{sec:intro}

Proteins are macromolecules that mediate the fundamental processes of all living organisms.
For decades, people are seeking to design novel proteins with desired properties~\citep{huang2016coming}, a problem known as \textit{de novo protein design}.
Nevertheless, the problem is very challenging due to the tremendous search space of both sequence and structure, and the most well-established approaches still rely on hand-crafted energy functions and heuristic sampling algorithms~\citep{leaver2013scientific, alford2017rosetta}, which are prone to arriving at suboptimal solutions and are computationally intensive and time-consuming.

Recently, machine learning approaches have demonstrated impressive performance on different aspects of protein design, and significant progress has been made~\citep{gao2020deep}.
Most approaches use deep generative models
to design protein sequences based on corresponding structures~\citep{ingraham2019generative, jing2021gvp, hsu2022learning}.
Despite their great potential for protein design, the structures of proteins to be engineered are often unknown~\citep{fischman2018computational}, which hinders the application of these methods.
Therefore, efforts have been made to develop models that \textit{co-design} the sequence and structure of proteins~\citep{anishchenko2021novo, wang2021deep}.
As a pioneering work, \cite{jin2021iterative} propose an autoregressive model that co-designs the Complementarity Determining Regions (CDRs) sequence and structure of antibodies based on iterative refinement of protein structures, which spurs a lot of follow-up works~\citep{luo2022antigen, kong2022conditional}.
Nevertheless, 
these approaches are tailored for antibodies and their effectiveness remains unclear on proteins with arbitrary domain topologies~\citep{anand2022protein}.
In addition, they often suffer from high inference costs due to autoregressive sampling or annealed diffusion sampling~\citep{song2019generative, luo2022antigen}.
Very recently, \cite{anand2022protein} propose another diffusion-based generative model~\citep{ho2020denoising} for general protein sequence-structure co-design, where they adopt three diffusion models to generate structures, sequences, and rotamers of proteins in sequential order.
Although applicable to proteins of all topologies, such a sequential generation strategy fails to cross-condition on sequence and structure, which might lead to inconsistent proteins.
Besides, the inference process is also expensive due to the use of three separate diffusion processes.

\begin{figure}[t]
\vspace{-5pt}
\centering
    \includegraphics[width=\textwidth]{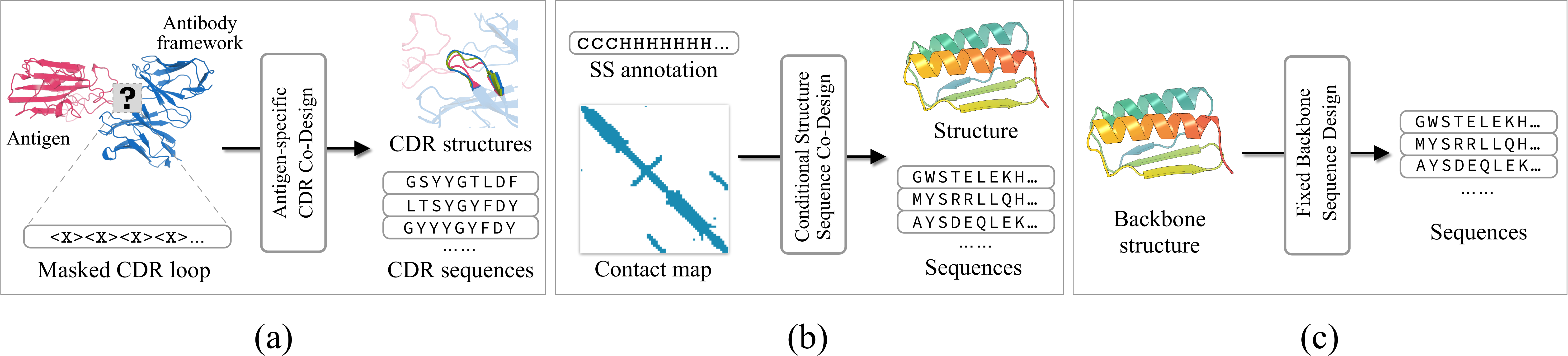}
    \vspace{-5pt}
    \caption{
    Illustration of three protein design tasks with different context features.
    (a) Antigen-specific CDR co-design given structure and sequence of antibody framework and the binding antigen.
    (b) Protein sequence-structure co-design conditioned on secondary structure (SS) annotation and binary contact features.
    (c) Fixed backbone sequence design conditioned on given backbone structures.
    }
\label{fig:intro}
\vspace{-5pt}
\end{figure}

To address the aforementioned issues, in this paper, we propose a new method capable of \underline{prot}ein \underline{se}quence-structure \underline{e}quivariant co-\underline{d}esign called \method.
Specifically, we formulate the co-design task as a translation problem in the joint sequence-structure space based on context features.
Here the context features represent prior knowledge encoding constraints that biologists want to impose on the protein to be designed~\citep{dou2018novo, shen2018novo}.
As an illustration, we present three protein design tasks with different given context features in Figure~\ref{fig:intro}.
Our \method consists of a trigonometry-aware encoder that infers geometrical constraints and prior knowledge for protein design from context features, and a novel roto-translation equivariant decoder that iteratively translates proteins into desired states in an end-to-end and equivariant manner.
The equivariance property with respect to protein structures during the whole process is guaranteed by predicting structure updates in local frames based on invariant representations, and then transforming them into global frames using change of basis operation.
It is worth mentioning that \method updates sequence and structure of all residues in an one-shot manner, leading to a much more efficient inference process.
In contrast to previous method that first generates structure and then generates sequence and rotamers, we allow the model to cross-condition on sequence and structure, and encourage the maximal information flow among context features, sequences, and structures, which ensure the fidelity of generated proteins.

We conduct extensive experiments on the Structural Antibody Database (SAbDab)~\citep{dunbar2014sabdab} as well as two protein design benchmark data sets curated from CATH~\citep{orengo1997cath}, and compare \method against previous state-of-the-art methods on multiple tasks, ranging from antigen-specific antibody CDR design to context-conditioned protein design and fixed backbone protein design.
Numerical results show that our method significantly outperforms previous baselines and can generate high fidelity proteins 
in terms of both sequence and structure, while running orders of magnitude faster than sampling-based methods.
As a proof of concept, we also show by cases that \method is able to perform \textit{de novo} protein design with new folds.

\section{Related Work}
\label{sec:related}

\textbf{Protein Design.}
The most well-established approaches on protein design mainly rely on handcrafted energy functions to iteratively search low-energy protein sequences and conformations with heuristic sampling algorithms~\citep{leaver2013scientific, alford2017rosetta, tischer2020design}.
Nevertheless, these conventional methods are computationally intensive, and are prone to arriving at local optimum due to the complicated energy landscape.
Recent advances in deep generative models open the door to data-driven approaches, and a variety of models have been proposed to generate protein sequences~\citep{rives2021biological, shin2021protein, ferruz2022protgpt2} or backbone structures~\citep{anand2018generative, eguchi2022igvae, trippe2022diffusion}.
To have fine-grain control over designed proteins, methods are developed to predict sequences that can fold into given backbone structures~\citep{ingraham2019generative, jing2021gvp, anand2022learned, dauparas2022robust}, a.k.a. fixed backbone design, which achieve promising results but require the desired protein structure to be known \textit{a priori}.

Recently, a class of approaches that generate both protein sequence and structure by network hallucination have emerged~\citep{anishchenko2021novo, wang2021deep}, which carry out thousands of gradient descent steps in sequence space to optimize loss functions calculated by pre-trained protein structure prediction models~\citep{yang2020improved, jumper2021highly}.
However, the quality of designed proteins usually relies on the accuracy of structure prediction models, and is sensitive to different random startings.
On the other hand, attempts have been made to co-design CDR sequence and structure of antibodies using either autoregressive models~\citep{saka2021antibody, jin2021iterative} or diffusion models~\citep{luo2022antigen}.
Nevertheless, they are restricted to proteins with specific domain topologies and often suffer from the time-consuming Monte Carlo sampling process.
Going beyond antibodies, \cite{anand2022protein} adopt three separate diffusion models to generate sequences, structures, and rotamers of proteins sequentially.
Such a method is inefficient and fails to cross-condition on both protein sequence and structure.
Our model also seeks to co-design protein sequence and structure, but is able to cross-condition on sequence and structure, while being much more efficient.

\textbf{3D Structure Prediction.}
Our work is also related to approaches that perform 3D structure prediction by iteratively translating structures in three dimensional space equivariantly~\citep{shi2021learning, luo2021diffusion, hoogeboom2022equivariant, xu2022geodiff, zhu2022direct}.
However, these methods represent structures as either molecular graphs or point clouds, and are not applicable to protein structures.
On the other hand, protein folding models~\citep{jumper2021highly, baek2021accurate} that perform protein structure prediction require complete protein sequences as well as their Multiple Sequence Alignments (MSAs) as input, and cannot co-design protein sequence and structure directly.


\section{Method}
\label{sec:method}

\subsection{Preliminaries}
\textbf{Notations.}
Proteins are macromolecules that can be viewed as chains of amino acids (residues) connected by peptide bonds. 
In this paper, an amino acid can be represented by its type $s_i \in \{1, \cdots 20\}$, C$_\alpha$ coordinates $\vx_i \in \sR^3$, and the frame orientation $\mO_i \in \mathrm{SO(3)}$, where $i \in \{1, \cdots N\}$ and $N$ is the number of residues in a protein.
The $\vx_i$ and $\mO_i$ form a canonical orientation frame with respect to the N, C and C$_\beta$ atoms, from which the backbone atom positions can be derived.
We denote the one-hot encoding of the residue type as $\vs_i = \operatorname{onehot}(s_i)$.
In the protein sequence and structure co-design task, researchers often provide context features as input to encourage designed proteins to have desired structural properties. 
These context features can either be single (residue) features $\vm_i \in \sR^{c_m}$ (e.g., amino acid secondary structure annotations) or pair features $\vz_{ij} \in \sR^{c_z}$ (e.g., binary contact features between residues).
With the above notations, a protein with $N$ residues can be compactly denoted as $\gP = \{(\vs_i, \vx_i, \mO_i) \}_{i=1}^N$.
The context features known \textit{a priori} can be denoted as $\{\vm_i\} \in \sR^{N \times {c_m}}$ and $\{\vz_{ij}\} \in \sR^ {N \times N \times {c_z}}$.

\textbf{Problem Formulation.}
Given a set of context features $\{\vm_i\} \in \sR^{N \times {c_m}}$ and $\{\vz_{ij}\} \in \sR^ {N \times N \times {c_z}}$, the task of \textit{protein sequence and structure co-design} is the joint generation of residue types and 3D conformations of a protein with $N$ residues, i.e., the conditional generation of $\gP = \{(\vs_i, \vx_i, \mO_i)\}_{i=1}^N$ based on $\{\vm_i\}$ and $\{\vz_{ij}\}$.
Note that context features vary from setting to setting. For example, in antibody CDR design~\citep{jin2021iterative, luo2022antigen}, they are derived from antibody framework and binding antigen structures with CDR region masked, while in full protein design~\citep{anand2022protein}, they can be secondary structure annotations and residue-residue contact features.

\begin{figure*}[t]
	\centering
	\vspace{-10pt}
    \includegraphics[width=1.0\linewidth]{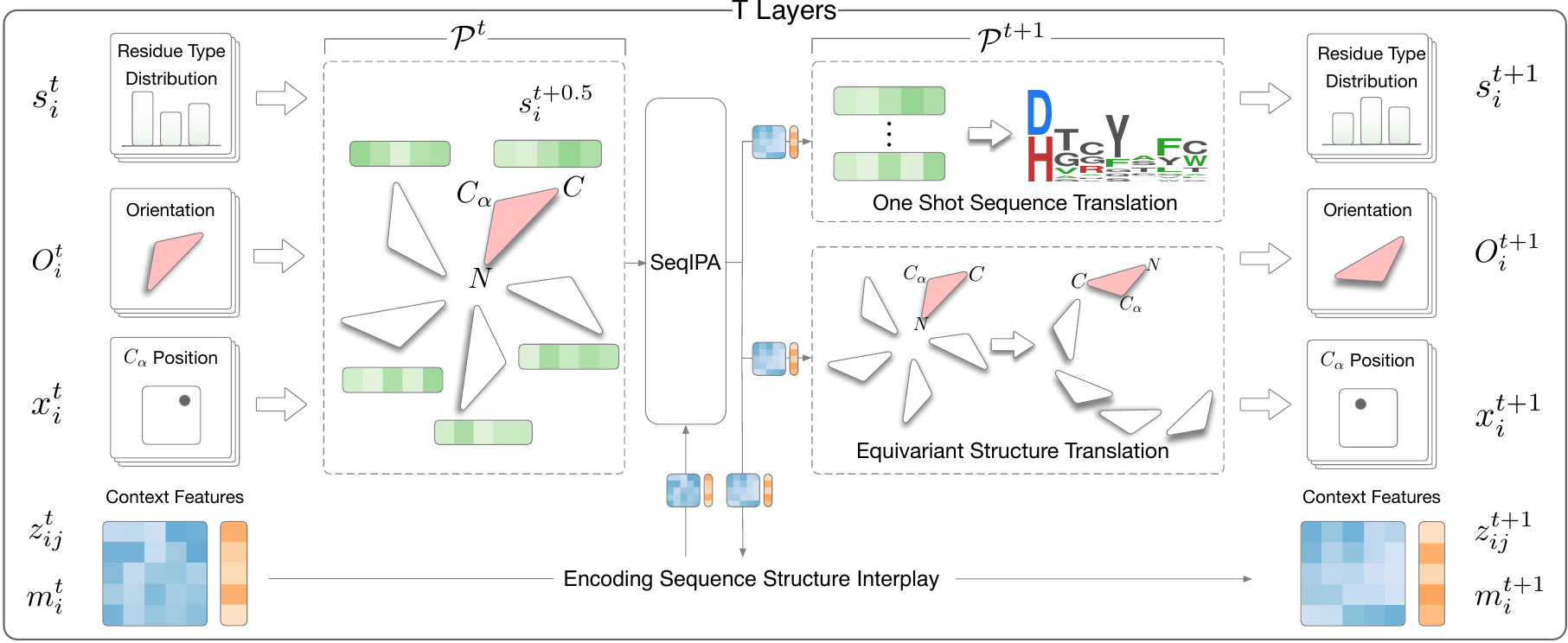}
    \vspace{-10pt}
    \caption{
    Illustration of the joint sequence-structure translation process.
    In each translation layer, the network first captures the interactions of the current protein state and context features via $\operatorname{SeqIPA}$, and then translates the protein sequence and structure into the next state equivariantly.
    }
    \label{fig:main}
    \vspace{-10pt}
\end{figure*}

\textbf{Overview.}
In this paper, we formulate protein sequence and structure co-design as an equivariant translation problem in the joint sequence-structure space.
Specifically, we develop a trigonometry-aware context encoder to first reason geometrical constraints encoded in context features.
Based on the updated context features, protein sequence and structure are jointly generated in an iterative manner by a novel roto-translation equivariant decoder, starting from randomly initialized structures and residue types (illustrated in Figure~\ref{fig:main}).
To model interactions between sequence and structure during decoding, we allow information to flow between context features, structures, and residue types in each translation step.
In this way, the generated sequence and structure are ensured to be consistent.
The pseudo-code of the whole framework can be found in Algorithm~\ref{alg:method}.
The rest of this section is organized as follows:
Section~\ref{subsec:encoder} introduces the trigonometry-aware context encoder.
Section~\ref{subsec:decoder} elaborates the iterative joint sequence-structure decoder and training objectives.



\subsection{Trigonometry-aware context encoder}
\label{subsec:encoder}
Given single features and pair features as input, the goal of the context encoder is to capture the interactions between different context features and infer encoded constraints for the following protein sequence-structure co-design. 
We first embed single features and pair features into $c$ dimensional space using Multiple Layer Perceptrons (MLPs). We then adopt a stack of $L$ trigonometry-aware update layer to propagate information between single features and pair features. 

We represent the updated single features and pair features at $l^{\mathrm{th}}$ layer as $\{\vm_i^l\}$ and $\{\vz_{ij}^l\}$, respectively.
At each layer, the single features are first updated using a variant of multi-head self-attention (denoted as $\operatorname{MHA}$)~\citep{vaswani2017attention}, with pair features serving as additional input to bias the attention map.
Similar to~\cite{jumper2021highly}, the pair feature $\vz_{ij}^l$ are then updated by the linear projection of outer product of single features $\vm_i^{l+1}$ and $\vm_j^{l+1}$:
\begin{align}
\label{eq:res_update}
\vm_i^{l+1}
&= \operatorname{MHA}(\{\vm_i^l\}, \{\vz_{ij}^l\}), \\
\label{eq:res2pair}
\vz_{ij}^{l+0.5} 
&= \vz_{ij}^{l} + \operatorname{Linear}(\vm_i^{l+1} \otimes \vm_j^{l+1}),
\end{align}
where $\otimes$ is the outer product operation.
Notably, we enable the information to flow between single features and pair features to better model interactions between context features. 

Since pair features (e.g., contact map and distance map) are usually related with Euclidean distances and dihedral angles between residues, inspired by AlphaFold 2~\citep{jumper2021highly}, we adopt two trigonometry-aware operations (Eq.~\ref{eq:triangle_mul} and Eq.~\ref{eq:triangle_att}) in each layer to maintain geometric consistency and encourage pair features to satisfy the triangle inequality. Formally, we have:
\begin{align}
\label{eq:pair_update}
\hat{\va}_{ij}, \hat{\vb}_{ij}
&= \sigma(\vz_{ij}^{l+0.5}) \odot \operatorname{Linear}(\vz_{ij}^{l+0.5}), 
\quad \vq_{ij}, \vk_{ij}, \vv_{ij}, \vb_{ij}
= \operatorname{Linear}(\vz_{ij}^{l+0.75}), \\
\label{eq:triangle_mul}
\vz_{ij}^{l+0.75}
&= \vz_{ij}^{l+0.5} + \sigma(\vz_{ij}^{l+0.5}) \odot \operatorname{Linear}\big(\sum\limits_k \hat{\va}_{ik}\odot \hat{\vb}_{jk} + \hat{\va}_{ki}\odot \hat{\vb}_{kj}\big), \\
\label{eq:triangle_att}
\vz_{ij}^{l+1}
&= \vz_{ij}^{l+0.75} + \sigma(\vz_{ij}^{l+0.75}) \odot \sum\limits_k (\alpha_{ijk} \vv_{ik} + \alpha_{ijk} \vv_{kj}),
\end{align}
where $\sigma(\cdot) = \operatorname{sigmoid}(\operatorname{Linear}(\cdot))$, and $\alpha_{ijk}=\operatorname{softmax}_k \big( \frac{1}{\sqrt{c}}\vq_{ij}^{\top}(\vk_{ik} + \vk_{kj}) + \vb_{jk} + \vb_{ki} \big)$ is the attention score of a novel trigonometry-aware attention. Intuitively, in our trigonometry-aware attention, the pair feature $\vz_{ij}$ is updated with neighboring features $\vz_{ik}$ and $\vz_{kj}$, by enumerating all possible $k$ that form a triangle $\Delta_{ijk}$ in terms of residues. Different from \cite{jumper2021highly}, we tie the attention score $\alpha_{ijk}$ within each triangle $\Delta_{ijk}$ to reduce the computational burden, while keeping the whole network sensitive to triangular interactions among residues. After $L$ rounds of feature propagation, the updated single features $\{\vm_i^L\}$ and pair features $\{\vz_{ij}^L\}$ serve as inputs to the decoder for joint protein sequence-structure design.

\subsection{Joint sequence-structure decoder}
\label{subsec:decoder}
In this section, we describe the proposed joint sequence-structure decoder, with the goal of iteratively translating protein sequence and structure into desired states from scratch based on context features.
Simply parameterizing the decoder using two neural networks that generate sequence and structure separately is problematic, as we need to ensure the consistency between generated sequence and structure, i.e., the sequence folds to the structure. 
Meanwhile, we require our decoder to be roto-translation equivariant~\citep{kohler20eqflow, jing2021gvp} with respect to all protein structures during the decoding.
To this end, we develop a novel roto-translation equivariant network composed of $T$ consecutive translation layers with weight tying~\citep{dehghani2018universal, bai2019deep}.
In each layer, we update context features, residue structures, and residue types
interdependently, by allowing information to propagate among them.
It is worth mentioning that residue types and residue structures of all amino acids are updated in one shot in each translation step, distinct from previous work that generates them autoregressively, which significantly accelerates the decoding procedure. 



We represent the updated residue types and residue structures at $t^{\mathrm{th}}$ layer as $\gP^t = \{(\vs_i^t, \vx_i^t, \mO_i^t) \}_{i=1}^N$. Abusing the notation a little bit, we denote updated context features at $t^{\mathrm{th}}$ layer as $\{\vm_i^t\}$ and $\{\vz_{ij}^t\}$.
Specifically, $\{\vm_i^0\}$ and $\{\vz_{ij}^0\}$ are the invariant output of context encoder introduced in Section~\ref{subsec:encoder}. For each amino acid, we initialize the residue type as the uniform distribution over 20 amino acid types, i.e., $\vs_i^0 = \frac{1}{20} \cdot \bm{1}$, C$_\alpha$ coordinates as origin in global frame, i.e., $\vx_i^0 = (0, 0, 0)$, and frame orientation as identity rotation, i.e., $\mO_i^0 = \mI_{\bm{3}}$. 
We elaborate each translation layer next.

\textbf{Encoding Sequence-Structure Interplay.}
In each translation layer, we start by embedding current residue types into $c$ dimensional space with a feedforward network $\operatorname{MLP_e}: \sR^{20} \rightarrow \sR^c, \vs_i^{t+0.5} = \operatorname{MLP_e}(\vs_i^t)$.
We then adopt a variant of Invariant Point Attention (IPA)~\citep{jumper2021highly} called $\operatorname{SeqIPA}$ to capture the interplay of residue types, residue structures and context features, integrating them all altogether into updated context features. Such a practice is often favored in literature~\citep{anand2022protein, luo2022antigen, tubiana2022scannet} as it is aware of the orientation of each residue frame while being roto-translation invariant to input and output features.
Distinct from vanilla IPA, our $\operatorname{SeqIPA}$ takes residue types as the additional input to bias the attention map and steer the representation of the whole protein generated so far:
\begin{align}
    \label{eq:seqipa}
    \vm_i^{t+1}, \vz_{ij}^{t+1} = \operatorname{SeqIPA}(\{\vm_i^{t}\}, \{\vz_{ij}^{t}\}, \{\vs_i^{t+0.5}\}, \{\vx_i^t\}, \{\mO_i^t\}).
\end{align}
Note that $\operatorname{SeqIPA}$ is orientation-aware with respect to residue structures and roto-translation invariant to all other representations. We refer readers to Appendix~\ref{subsec:suppl_seqipa} for more details about the $\operatorname{SeqIPA}$.

\textbf{Equivariant Structure Translation.}
Given updated context features, the protein structure is translated towards the next state by updating $\{\vx_i^t\}$ and $\{\mO_i^t\}$.
To update C$_\alpha$ positions $\{\vx_i^t\}$, we first predict the change of coordinates (denoted as $\{\hat{\vx}_i^t\}$) within each local residue frame specified by $\{\mO_i^t\}$.
We then perform a change of basis using $\{\mO_i^t\}$ to transform $\{\hat{\vx}_i^t\}$ from local frame into global frame~\citep{kofinas2021roto, hsu2022learning} to derive equivariant deviation of C$_\alpha$ positions (Eq.~\ref{eq:ca_update}). Intuitively, the deviation of C$_\alpha$ positions rotate accordingly when residue frames rotate, which guarantees the equivariance of the C$_\alpha$ translation step.

The update for orientation frame $\mO_i^t$ is computed by predicting a unit quaternion vector~\citep{jia2008quaternions} with a feedforward network, which is then converted to a rotation matrix $\hat{\mO}_i^t$, and left-multiplied by $\mO_i^t$ to rotate the residue frame (Eq.~\ref{eq:frame_update}).
We adopt the unit quaternion here because it is a more concise representation of a rotation in 3D than a rotation matrix.
Since the predicted unit quaternion is an invariant vector, the predicted rotation matrix $\hat{\mO}_i^t$ is also invariant.
Therefore, the translation step of the residue frame is equivariant due to the multiplication of rotation matrices.
We summarize the whole equivariant structure translation step as follows:
\begin{align}
\label{eq:ca_update}
\hat{\vx}_i^t
&= \operatorname{MLP_x}(\vm_i^{t+1}, \vm_i^{0}),
\quad \vx_i^{t+1} = \vx_i^{t} + \mO_i^t \hat{\vx}_i^t, \\
\label{eq:frame_update}
\hat{\mO}_i^t
&= \operatorname{convert}\big(\operatorname{MLP_o}(\vm_i^{t+1}, \vm_i^{0})\big),
\quad \mO_i^{t+1} = \mO_i^t \hat{\mO}_i^t,
\end{align}
where $\operatorname{convert}$ is a function that converts a quaternion to a rotation matrix.

\textbf{One Shot Sequence Translation.}
The residue type of all amino acids are updated in one shot based on updated context features and current residue types in each translation step.
Specifically, we use a feedforward network $\operatorname{MLP_s}$ to predict residue type distributions over 20 amino acid types for the next iteration:
\begin{align}
\label{eq:seq_update}
\vs_i^{t+1}
&= \operatorname{softmax} \big(\lambda \cdot \operatorname{MLP_s}(\vm_i^{t+1}, \vm_i^{0}, \vs_i^{t+0.5})\big ),
\end{align}
where $\lambda$ is a hyper-parameter controlling the temperature of the distribution.
See Appendix~\ref{subsec:hyper_parameters} for discussions on hyper-parameters and the equivariance property of sequence and structure translation.

Summarizing the above, at $(t+1)^{\mathrm{th}}$ layer, the decoder takes $\gP^t = \{(\vs_i^t, \vx_i^t, \mO_i^t)\}_{i=1}^N$ as the input and computes $\gP^{t+1} = \{(\vs_i^{t+1}, \vx_i^{t+1}, \mO_i^{t+1})\}_{i=1}^N$ as the output.
Based on $\gP^t$, we can efficiently reconstruct full backbone atom positions according to their averaged relative positions with respect to $C_\alpha$ recorded in literature~\citep{engh2012structure, jumper2021highly}.
It is worth mentioning that distinct from previous works that can only generate backbone atom positions~\citep{jin2021iterative, kong2022conditional, luo2022antigen}, our model is capable of full atom position generation by attaching the corresponding sidechain to each residue.
In specific, we can enable the decoder to generate four additional torsion angles ($\chi_1, \chi_2, \chi_3, \chi_4$)~\citep{mcpartlon2022end} that specify the geometry of sidechain atoms and reconstruct sidechain atom positions together with backbone atom positions.

\textbf{Training Objective.}
We denote the reconstructed full backbone atom positions at $t^{\mathrm{th}}$ layer as $\{\vx_{ij}^{t}\}$, where $j \in \{1,2,3\}$ is the index of backbone atoms $(N, C_\alpha, C)$. We use the superscript $\mathrm{true}$ to denote the ground truth value for simplicity.
The whole network can be jointly optimized, by defining an invariant cross-entropy loss $\ell_{\mathrm{ce}}$ over type distributions and an invariant frame align loss~\citep{jumper2021highly} over full backbone atom positions at each translation layer:
\begin{align}
\label{eq:type_loss}
\gL_{\mathrm{type}} 
&= \frac{1}{T\cdot N}\sum\limits_{t=1}^T\sum\limits_{i=1}^N \ell_{\mathrm{ce}}(\vs_i^t, \vs_i^{\mathrm{true}}), \\
\label{eq:pos_loss}
\gL_{\mathrm{pos}}
&= \frac{1}{T\cdot N \cdot 3N}\sum\limits_{t=1}^T\sum\limits_{k=1}^N\sum\limits_{i,j}
\Vert \rho_k^t (\vx_{ij}^t) - \rho_k^{\mathrm{true}}(\vx_{ij}^{\mathrm{true}})\Vert_2,
\end{align}
where $\rho_k^t(\vx_{ij}^t) = (\mO_k^t)^{-1}(\vx_{ij}^t -\vx_k^t)$ transforms the backbone coordinates from global coordinate system into local frame of $k^{\mathrm{th}}$ residue frame at step $t$ (specified by $\vx_k^t$ and $\mO_k^t$), and so does $\rho_k^{\mathrm{true}}$.
The loss defined in Eq.~\ref{eq:pos_loss} essentially calculates the discrepancy of all backbone positions between the prediction and the ground truth, \textit{by aligning each residue frame one by one}.
As all backbone coordinates are transformed into local frames, the loss will stay invariant when two protein structures differ by an arbitrary rotation and an arbitrary translation. Combining Eq.~\ref{eq:type_loss} and Eq.~\ref{eq:pos_loss}, the final training objective is $\gL = \gL_{\mathrm{type}} + \gL_{\mathrm{pos}}$.

\section{Experiments}
\label{sec:experiment}

Following previous works~\citep{jin2021iterative, jing2021gvp, anand2022protein}, we conduct extensive experiments and evaluate the proposed \method on
the following three tasks:
\textbf{Antibody CDR Co-Design} (Section~\ref{sec:ab_design}), \textbf{Protein Sequence-Structure Co-Design} (Section~\ref{sec:prot_design}), and \textbf{Fixed Backbone Sequence Design} (Section~\ref{sec:fixbb_design}).
We also show cases where \method successfully conducts \textit{de novo} protein sequence design with new folds in Section~\ref{sec:case}.
We describe all experimental setups and results in task-specific sections.

\subsection{Antibody CDR Co-Design}
\label{sec:ab_design}

\textbf{Setup.}
The first task is to design CDR sequence and structure of antibodies, where the context features are amino acid types as well as inter-residue distances derived from antibody-antigen complexes with CDRs removed.
The initial protein structure $\gP^0$ is set as the complex structure except for CDRs, which are randomly initialized.
We retrieve antibody-antigen complexes in Apr. 2022 from Structural Antibody Database (SAbDab)~\citep{dunbar2014sabdab}, and remove incomplete or redundant complexes, resulting in a subset containing 2,900 complex structures.
Following \cite{jin2021iterative}, we focus on the design of heavy chain CDRs and curate three data splits for each type of CDRs (denoted as H1, H2, H3) by clustering corresponding CDR sequences via MMseqs2~\citep{steinegger2017mmseqs2} with 40\% sequence identity.
In total, there are 641, 963, and 1646 clusters for CDR H1, H2, and H3.
The clusters are then divided into training, validation, and test set with a ratio of 8:1:1.

We compare \method with the following three baselines. RosettaAntibodyDesign (\textbf{RAbD})~\citep{adolf2018rosettaantibodydesign} is a physics-based antibody design software. \textbf{GNN} is an autoregressive model that co-designs sequence and structure similar to \cite{jin2021iterative}.
We note that we can not directly compare \method with \cite{jin2021iterative} as the setting is different.
\textbf{Diffusion}~\citep{luo2022antigen} is a diffusion-based method that achieves state-of-the-art performance on antibody design recently.
Following previous works~\citep{jin2021iterative,luo2022antigen}, we use three metrics to evaluate the quality of designed CDRs:
(1) Perplexity (\textbf{PPL}) measures the inverse likelihood of native sequences in the predicted sequence distribution, and a lower PPL stands for the higher likelihood.
Note that PPL should be calculated \textit{for each sequence first and then averaged over the test set.}
For methods that do not define joint distributions over sequences strictly (e.g., \diffusion), we use the final sequence distribution to calculate the approximated PPL.
(2) \textbf{RMSD} is the Root of Mean Squared Deviation of C$_\alpha$ between generated CDRs and ground-truth CDRs \textit{with antibody frameworks aligned}~\citep{ruffolo2022antibody}. A lower RMSD stands for a smaller discrepancy compared to native CDRs, which indicates a better CDR structure to bind the antigen.
(3) Amino acid recovery rate (\textbf{AAR}) is the sequence identity between generated and ground-truth CDRs.

\vspace{-5pt}
\begin{table}[htbp]
    \centering
    \vspace{-5pt}
    \caption{
        PPL, RMSD and AAR of different approaches on the antibody CDR co-design task. ($\uparrow$): the higher the better. ($\downarrow$): the lower the better.
    }
    \label{tab:antibody}
    \vspace{2pt}  

    \begin{adjustbox}{width=1.0\linewidth}
        \begin{tabular}{l | ccc | ccc | ccc}
\toprule
& \multicolumn{3}{c|}{PPL ($\downarrow$)} 
& \multicolumn{3}{c|}{RMSD (\AA, $\downarrow$)} 
& \multicolumn{3}{c}{AAR (\%, $\uparrow$)} \\

CDR 
& H1 & H2 & H3
& H1 & H2 & H3
& H1 & H2 & H3 \\
\midrule
\rosetta
& --- & --- & ---
&3.06 (.33) & 2.95 (.23) & 5.58 (.32)
& 27.36 (1.89) & 47.11 (1.74) & 22.91 (2.18) \\
\gnn
&5.74 (.45) &7.64 (.53) &13.91 (.39)
&2.73 (.06) &3.14 (.13) &4.06 (.14)
&61.49 (.27) &52.40 (.08) &29.83 (.32) \\
\diffusion
& 4.72 (.09) & 6.28 (.10) & 12.27 (.11) 
& 1.94 (.12) & 1.92 (.37) & 3.95 (.01)
& 66.38 (.95) & 55.61 (.03) & 31.22 (.79) \\
\midrule
\textbf{\method}
& \textbf{4.43 (.40)}  & \textbf{5.94 (.17)}  & \textbf{10.88 (.38)}
& \textbf{1.24 (.03)}  & \textbf{1.11 (.04)}  & \textbf{3.19 (.03)}
& \textbf{70.22 (.98)}  & \textbf{63.53 (.85)}  & \textbf{39.27 (.77)}
\\
\bottomrule
\end{tabular}

    \end{adjustbox}
    \vspace{-5pt}
\end{table}

\begin{figure}[thbp]
\centering

    \includegraphics[width=1.0\linewidth]{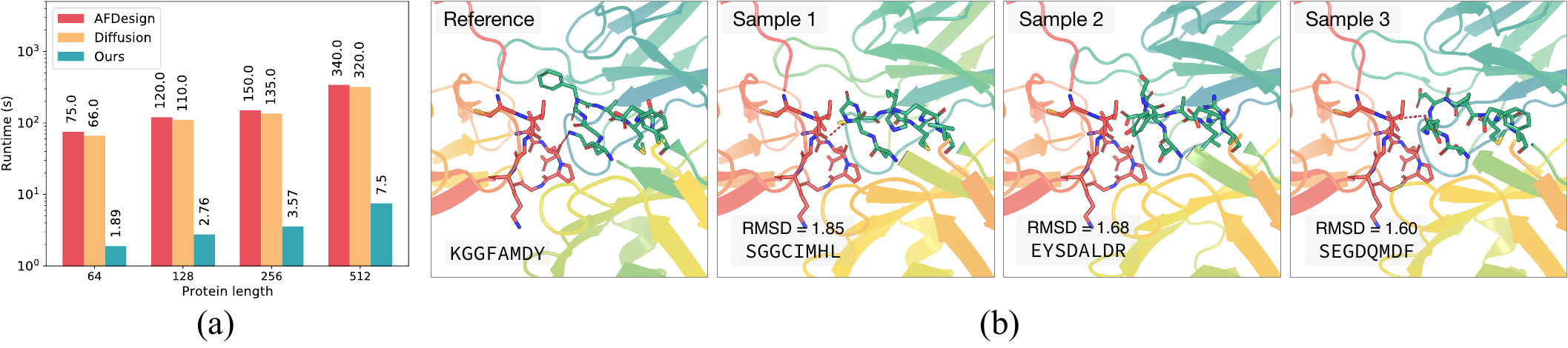}
    \vspace{-10pt}
    \vspace{-7pt}    
\caption{
    (a) Average runtime of three approaches on proteins of different sizes. Our \method runs orders of magnitude faster than the gradient-based method \afdesign and the diffusion-based method \diffusion.
    (b) Examples of CDR-H3 sequence and structure co-designed by our method. The PDB ID of the reference complex is \href{https://www.rcsb.org/structure/6FLA}{6FLA}, and the antigen is Dengue Virus.
    The sidechain interactions between the antigen (red) and the antibody (green) are highlighted.
}
\label{fig:runtime_antibody}
\vspace{-10pt}
\end{figure}

\textbf{Results.}
We notice that scripts used to calcuate the above metrics are inconsistent in previous works.
For a fair comparison, we implement all the baselines and run each model three times with different random seeds.
Following previous works~\citep{jin2021iterative, luo2022antigen}, the length of the CDR is set to be identical to the length of the ground-truth CDR for simplicity, and we sample 100 candidates with the lowest perplexity for each CDR for machine learning-based methods.
We report the mean and standard deviation of the above metrics on the test set in Table~\ref{tab:antibody}.
Numerical results indicate that \method consistently outperforms previous state-of-the-art baselines by a clear margin on all three metrics for each type of CDRs, which confirms \method's ability to co-design antibodies conditioned on existing binding structures.
In particular, as an energy-based method that performs graft antibody design, the performance of \rosetta is inferior to data-driven approaches.
\diffusion outperforms \gnn as it is equivariant and models the orientation of residues similar to our model, but its performance still falls short of ours.
It is worth mentioning that the performance of all models on CDR H3 is worse than that on the other two CDR types, as CDR H3 is the most diverse region in an antibody that is critical to antigen binding.
We present generated samples on CDR-H3 sequence-structure co-design and their sidechain interactions with binding antigens in Figure~\ref{fig:runtime_antibody}(b).

\subsection{Protein Sequence and Structure Co-Design}
\label{sec:prot_design}

\textbf{Setup.}
This task evaluates models' capability to design protein sequence and structure conditioned on context features known \textit{a priori}.
The task is brought into attention by~\cite{anand2022protein} recently, but there is no rigorous benchmark in the field of machine learning to the best of our knowledge.
To this end, we collect 31,877 protein structures from the non-redundant set of CATH S40~\citep{orengo1997cath}, calculate amino acid secondary structure by DSSP~\citep{kabsch1983dictionary}, and split the data set into training, validation, and test set at the topology level with a 90/5/5 ratio.
We take secondary structure annotations as single features and binary contact matrices as pair features.\footnote{In this work, we say two residues are in contact if the distance between two C$_\alpha$ is within 8 \AA.}
This setting represents the situation where biologists already know the topology of desired proteins~\citep{dou2018novo, shen2018novo}, and want the model to design novel proteins without specifying structures.

For this task, we compare \method with the following two baselines that we believe best describe the current landscape of this task. \textbf{AFDesign}\footnote{\url{https://github.com/sokrypton/ColabDesign}} is a hallucination-based~\citep{wang2021deep} method that generates protein sequence and structure by iteratively performing gradient descent in sequence space guided by AlphaFold2. 
We adopt \afdesign as one of our baselines because AlphaFold2 is the state-of-the-art protein structure prediction model, and it outperforms other hallucination-based methods.
\textbf{Diffusion}~\citep{anand2022protein} is another diffusion-based model that generates protein structure, sequence, and rotamers sequentially, and it is not restricted to antibodies.
We use the \textbf{PPL}, \textbf{RMSD}, and \textbf{AAR} metrics introduced in Section~\ref{sec:ab_design} to evaluate the fidelity of designed proteins.

\begin{wrapfigure}[9]{R}{0.57\textwidth}
\vspace{-2.3em}
\begin{minipage}{0.57\textwidth}
\begin{table}[H]
    \centering
    \vspace{-7pt}    
    \caption{PPL, RMSD and AAR of different approaches on the protein co-design task. ($\uparrow$): the higher the better. ($\downarrow$): the lower the better.}
    \vspace{+1pt}
    \vspace{1pt}  
    \label{tab:protein}
    \resizebox{\columnwidth}{!}{
    \scalebox{1}{
        \begin{tabular}{c | ccc}
\toprule
Method & PPL ($\downarrow$) & RMSD (\AA, $\downarrow$) & AAR (\%, $\uparrow$) \\
\midrule
\afdesign &  --- &  3.47 (.11) & 12.05 (.28) \\
\diffusion & 11.63 (.08) & 2.33 (.37) & 24.97 (.32) \\
\midrule
\method & \bf 8.87 (.17) & \bf 1.29 (.02) & \bf 33.10 (.34) \\
\bottomrule
\end{tabular}

    }}
\end{table}
\end{minipage}

\end{wrapfigure}

\textbf{Results.}
Since \afdesign is incapble of handling discrete contact matrices, we set the loss function that guides the gradient descent as the distogram loss computed by a pre-trained AlphaFold2~\citep{jumper2021highly}.
For \diffusion, we implement it by ourselves as its source code is not released yet.
We run each model for three times with different random seeds and report the mean and standard deviation of the above metrics on the test set.
As shown in Table~\ref{tab:protein}, \method outperforms all the baselines on all three metrics significantly, which demonstrates its capability to generate proteins of high fidelity as regards both sequence and structure, and it is not limited to specific domain topologies.
The performance of \afdesign falls short of other methods as it relies on gradient descent to optimize the sequence, which is prone to getting stuck in the local optimum due to the rough landscape of the loss function.
The performance of \diffusion is also inferior to ours, as it generates protein using three separate diffusion models and fails to cross-condition on structure and sequence. In contrast, \method updates sequence and structure interdependently in an end-to-end manner.

To demonstrate the efficiency of our method, we test inference stage of different approaches using a single V100 GPU card on the same machine, and present average runtime of these methods on proteins of different sizes.
As indicated by Figure~\ref{fig:runtime_antibody}(a), our \method runs orders of magnitude faster than two baseline models on all four protein sizes, as both \afdesign and \diffusion rely on time-consuming Monte Carlo sampling steps.

\vspace{-5pt}
\begin{table}[htbp]
    \centering
    \vspace{-5pt}    
    \caption{
    PPL and AAR of different approaches on the fixed backbone sequence design task. ($\uparrow$): the higher the better. ($\downarrow$): the lower the better.
    Results of baselines are taken from~\cite{jing2021gvp}.
    }
    \vspace{2pt}  
    \label{tab:fixbb}
    \resizebox{0.85\columnwidth}{!}{
    \scalebox{1}{
        \begin{tabular}{c | ccc | ccc }
\toprule
\multirow{2}{*}{Method}
& \multicolumn{3}{c|}{PPL ($\downarrow$)} 
& \multicolumn{3}{c}{AAR (\%, $\uparrow$)} \\
& Short & Single-chain & All
& Short & Single-chain & All \\
\midrule
\esmif
& 8.94 & 8.67 & 6.70
& 27.3 & 28.3 & 36.5 \\
\structgnn
& 8.31 & 8.88 & 6.55
& 28.4 & 28.1 & 37.3 \\
\gvp
& \bf 7.10 & 7.44 & \bf 5.29
& 32.1 & 32.0 & 40.2 \\
\midrule
\method
& 7.32 & \bf 7.38 & 5.60
& \bf 34.8 & \bf 34.1 & \bf 43.8 \\
\bottomrule
\end{tabular}

    }}
    \vspace{-10pt}
\end{table}

\subsection{Fixed Backbone Sequence Design}
\label{sec:fixbb_design}

\textbf{Setup.}
The third task is to design protein sequences that can fold into given backbone structures, which is known as fixed backbone design.
In this task, context features are dihedral angles and inter-residue distances derived solely from backbone coordinates~\citep{jing2021gvp}, and protein structures are fixed as ground truth in the decoder.
We use the CATH 4.2 dataset curated by~\cite{ingraham2019generative}, and follow all experimental settings of~\cite{jing2021gvp} rigorously for a fair comparison, i.e., using the same data splits and evaluation settings according to their official implementation.
We compare \method with three baselines.
Specifically, \textbf{Structured GNN} is an improved version of~\cite{ingraham2019generative}. \textbf{GVP-GNN}~\citep{jing2021gvp} and \textbf{GVP-Transformer}~\citep{hsu2022learning} are state-of-the-art methods for fixed backbone design built upon Geometric Vector Perceptron (GVP) encoders~\citep{jing2021gvp}. Note that since we can not afford training models on AlphaFold2-predicted data sets~\citep{hsu2022learning} which requires hundreds of GPU days, we focus on the CATH data set for training.
We evaluate the performance of all methods using \textbf{PPL} and \textbf{AAR} as introduced in Section~\ref{sec:ab_design}, and drop the \textbf{RMSD} metric as structures are provided in this task.

\textbf{Results.}
Following~\cite{jing2021gvp}, we report the evaluation results on three test splits including the full CATH 4.2 test set, the short subset (100 or fewer residues) and the single-chain subset.
As shown in Table~\ref{tab:fixbb}, \method achieves competitive results against the state-of-the-art method \gvp in terms of the perplexity, and outperforms all baselines in terms of the amino acid recovery rate.
The results indicate that \method is a quite general protein design framework and is also superior in designing protein sequences conditioned on desired backbone structures.
Note that the state-of-the-art method \esmif is outperformed by \gvp when trained solely on CATH data set, which is consistent with the results reported in the original paper~\citep{hsu2022learning}.

\subsection{Case Study}
\label{sec:case}

\begin{figure*}[t]
	\centering
	\vspace{-10pt}
    \includegraphics[width=0.95\linewidth]{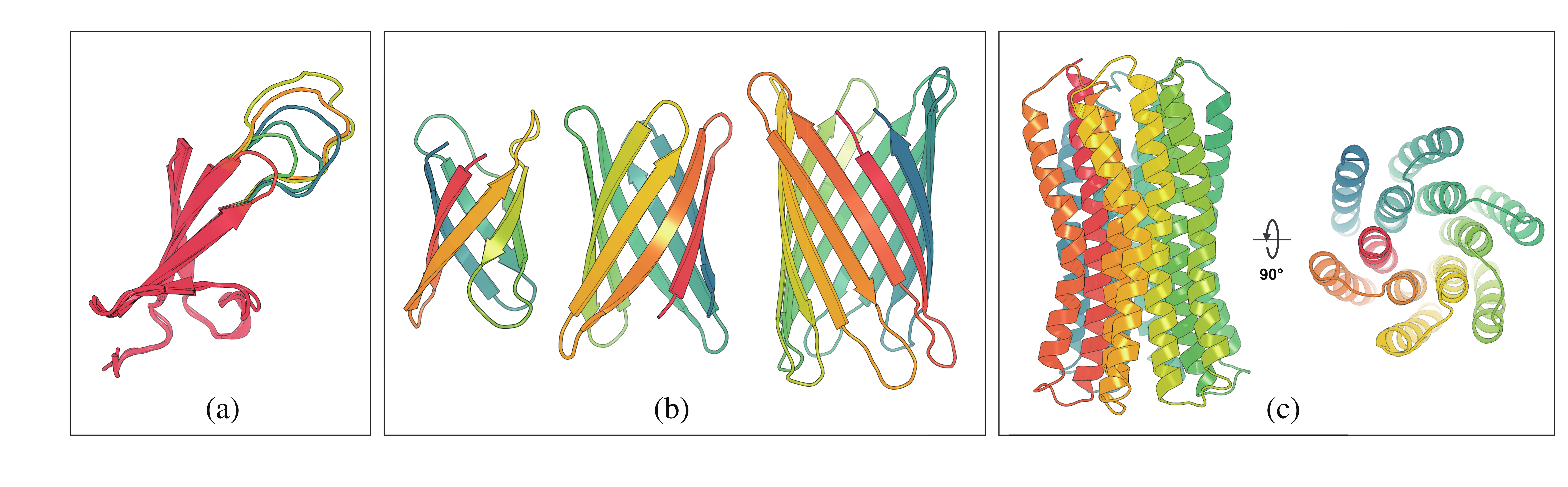}
    \vspace{-1pt}
    \caption{
    Example of novel proteins designed by \method.
    (a) Extending the loop of a native protein (marked in red).
    (b) Novel $\beta-$barrel design with different sizes.
    (c) Transmembrane protein complex design with a custom number of (twelve) $\alpha-$helices.
    }
    \label{fig:case_study}
    \vspace{-10pt}
\end{figure*}

So far, we have evaluated \method's capability to design proteins of high fidelity on multiple settings. However, it remains unclear whether the proposed model can go beyond the topologies of existing proteins. 
To get insight into the proposed method and as a proof of concept, we manually construct a set of secondary structure annotations and contact features from scratch, and ask the model trained in the second task to perform \textit{de novo} protein design based on the context features provided by us.
In Figure~\ref{fig:case_study}, we show that our \method succeeds in altering loop lengths of existing proteins, designing novel proteins with idealized topologies, and designing novel protein complexes with a custom number of secondary structures.
Notably, the designed structures are in close agreement with structures predicted by AlphaFold, taking the designed sequences as input.
We further perform protein sequence and structure search against all available databases using FoldSeek~\citep{van2022foldseek} and BLAST~\citep{altschul1990basic}, and find these synthetic proteins are dissimilar to existing proteins regarding both sequence and structure.
The case study serves as the first attempt to apply our model to \textit{de novo} protein design in a more realistic setting, which reveals the possibility of \method being a powerful tool for protein design in biological research.
We refer readers to Appendix~\ref{subsec:case_study} for all details about the case study.

\section{Conclusion and Future Work}
\label{sec:conclusion}

In this paper, we propose a novel principle for protein sequence and structure co-design called \method, which translates proteins in the joint sequence-structure space in an iterative and end-to-end manner.
\method is capable of capturing the interplay of sequence, structure, and context features during the translation, and owns a much more efficient inference process thanks to the one-shot translation strategy.
Extensive experiments over a wide range of protein design tasks show that \method outperforms previous state-of-the-art baselines by a large margin, confirming the superiority and generality of our method.
Further case studies on \textit{de novo} protein design demonstrate \method'{s} potential for more practical applications in biological research.
Future work includes extending \method to the scaffolding task~\citep{trippe2022diffusion} and adopting latent variables~\citep{kingma2013auto} to enable the context-free protein design.
\clearpage
\subsubsection*{Reproducibility Statement}
For the sake of reproducibility, the pseudo-code of \method, the parameterization of $\operatorname{SeqIPA}$, as well as hyper-parameters and implementation details are provided in Appendix~\ref{sec:suppl_model_details}.
All codes, datasets, and experimental environments will be released upon the acceptance of this work.
\subsubsection*{Acknowledgement}
We would like to thank all the reviewers for their insightful comments.
Jian Tang is supported by Twitter, Intel, the Natural Sciences and Engineering Research Council (NSERC) Discovery Grant, the Canada CIFAR AI Chair Program, Samsung Electronics Co., Ltd., Amazon Faculty Research Award, Tencent AI Lab Rhino-Bird Gift Fund, an NRC Collaborative R\&D Project (AI4D-CORE-06) as well as the IVADO Fundamental Research Project grant PRF-2019-3583139727.

\bibliography{reference}

\begin{thebibliography}{54}
\providecommand{\natexlab}[1]{#1}
\providecommand{\url}[1]{\texttt{#1}}
\expandafter\ifx\csname urlstyle\endcsname\relax
  \providecommand{\doi}[1]{doi: #1}\else
  \providecommand{\doi}{doi: \begingroup \urlstyle{rm}\Url}\fi

\bibitem[Adolf-Bryfogle et~al.(2018)Adolf-Bryfogle, Kalyuzhniy, Kubitz,
  Weitzner, Hu, Adachi, Schief, and
  Dunbrack~Jr]{adolf2018rosettaantibodydesign}
Jared Adolf-Bryfogle, Oleks Kalyuzhniy, Michael Kubitz, Brian~D Weitzner,
  Xiaozhen Hu, Yumiko Adachi, William~R Schief, and Roland~L Dunbrack~Jr.
\newblock Rosettaantibodydesign (rabd): A general framework for computational
  antibody design.
\newblock \emph{PLoS computational biology}, 14\penalty0 (4):\penalty0
  e1006112, 2018.

\bibitem[Alford et~al.(2017)Alford, Leaver-Fay, Jeliazkov, O’Meara, DiMaio,
  Park, Shapovalov, Renfrew, Mulligan, Kappel, et~al.]{alford2017rosetta}
Rebecca~F Alford, Andrew Leaver-Fay, Jeliazko~R Jeliazkov, Matthew~J O’Meara,
  Frank~P DiMaio, Hahnbeom Park, Maxim~V Shapovalov, P~Douglas Renfrew,
  Vikram~K Mulligan, Kalli Kappel, et~al.
\newblock The rosetta all-atom energy function for macromolecular modeling and
  design.
\newblock \emph{Journal of chemical theory and computation}, 13\penalty0
  (6):\penalty0 3031--3048, 2017.

\bibitem[Altschul et~al.(1990)Altschul, Gish, Miller, Myers, and
  Lipman]{altschul1990basic}
Stephen~F Altschul, Warren Gish, Webb Miller, Eugene~W Myers, and David~J
  Lipman.
\newblock Basic local alignment search tool.
\newblock \emph{Journal of molecular biology}, 215\penalty0 (3):\penalty0
  403--410, 1990.

\bibitem[Anand \& Achim(2022)Anand and Achim]{anand2022protein}
Namrata Anand and Tudor Achim.
\newblock Protein structure and sequence generation with equivariant denoising
  diffusion probabilistic models.
\newblock \emph{arXiv preprint arXiv:2205.15019}, 2022.

\bibitem[Anand \& Huang(2018)Anand and Huang]{anand2018generative}
Namrata Anand and Possu Huang.
\newblock Generative modeling for protein structures.
\newblock \emph{Advances in neural information processing systems}, 31, 2018.

\bibitem[Anand et~al.(2022)Anand, Eguchi, Mathews, Perez, Derry, Altman, and
  Huang]{anand2022learned}
Namrata Anand, Raphael Eguchi, Irimpan~I Mathews, Carla~P Perez, Alexander
  Derry, Russ~B Altman, and Po-Ssu Huang.
\newblock Protein sequence design with a learned potential.
\newblock \emph{Nature communications}, 13\penalty0 (1):\penalty0 1--11, 2022.

\bibitem[Anishchenko et~al.(2021)Anishchenko, Pellock, Chidyausiku, Ramelot,
  Ovchinnikov, Hao, Bafna, Norn, Kang, Bera, et~al.]{anishchenko2021novo}
Ivan Anishchenko, Samuel~J Pellock, Tamuka~M Chidyausiku, Theresa~A Ramelot,
  Sergey Ovchinnikov, Jingzhou Hao, Khushboo Bafna, Christoffer Norn, Alex
  Kang, Asim~K Bera, et~al.
\newblock De novo protein design by deep network hallucination.
\newblock \emph{Nature}, 600\penalty0 (7889):\penalty0 547--552, 2021.

\bibitem[Baek et~al.(2021)Baek, DiMaio, Anishchenko, Dauparas, Ovchinnikov,
  Lee, Wang, Cong, Kinch, Schaeffer, et~al.]{baek2021accurate}
Minkyung Baek, Frank DiMaio, Ivan Anishchenko, Justas Dauparas, Sergey
  Ovchinnikov, Gyu~Rie Lee, Jue Wang, Qian Cong, Lisa~N Kinch, R~Dustin
  Schaeffer, et~al.
\newblock Accurate prediction of protein structures and interactions using a
  three-track neural network.
\newblock \emph{Science}, 373\penalty0 (6557):\penalty0 871--876, 2021.

\bibitem[Bai et~al.(2019)Bai, Kolter, and Koltun]{bai2019deep}
Shaojie Bai, J~Zico Kolter, and Vladlen Koltun.
\newblock Deep equilibrium models.
\newblock \emph{Advances in Neural Information Processing Systems}, 32, 2019.

\bibitem[Dauparas et~al.(2022)Dauparas, Anishchenko, Bennett, Bai, Ragotte,
  Milles, Wicky, Courbet, de~Haas, Bethel, et~al.]{dauparas2022robust}
Justas Dauparas, Ivan Anishchenko, Nathaniel Bennett, Hua Bai, Robert~J
  Ragotte, Lukas~F Milles, Basile~IM Wicky, Alexis Courbet, Rob~J de~Haas,
  Neville Bethel, et~al.
\newblock Robust deep learning--based protein sequence design using
  proteinmpnn.
\newblock \emph{Science}, pp.\  eadd2187, 2022.

\bibitem[Dehghani et~al.(2018)Dehghani, Gouws, Vinyals, Uszkoreit, and
  Kaiser]{dehghani2018universal}
Mostafa Dehghani, Stephan Gouws, Oriol Vinyals, Jakob Uszkoreit, and {\L}ukasz
  Kaiser.
\newblock Universal transformers.
\newblock \emph{arXiv preprint arXiv:1807.03819}, 2018.

\bibitem[Dou et~al.(2018)Dou, Vorobieva, Sheffler, Doyle, Park, Bick, Mao,
  Foight, Lee, Gagnon, et~al.]{dou2018novo}
Jiayi Dou, Anastassia~A Vorobieva, William Sheffler, Lindsey~A Doyle, Hahnbeom
  Park, Matthew~J Bick, Binchen Mao, Glenna~W Foight, Min~Yen Lee, Lauren~A
  Gagnon, et~al.
\newblock De novo design of a fluorescence-activating $\beta$-barrel.
\newblock \emph{Nature}, 561\penalty0 (7724):\penalty0 485--491, 2018.

\bibitem[Dunbar et~al.(2014)Dunbar, Krawczyk, Leem, Baker, Fuchs, Georges, Shi,
  and Deane]{dunbar2014sabdab}
James Dunbar, Konrad Krawczyk, Jinwoo Leem, Terry Baker, Angelika Fuchs, Guy
  Georges, Jiye Shi, and Charlotte~M Deane.
\newblock Sabdab: the structural antibody database.
\newblock \emph{Nucleic acids research}, 42\penalty0 (D1):\penalty0
  D1140--D1146, 2014.

\bibitem[Eguchi et~al.(2022)Eguchi, Choe, and Huang]{eguchi2022igvae}
Raphael~R. Eguchi, Christian~A Choe, and Po-Ssu Huang.
\newblock Ig-vae: Generative modeling of protein structure by direct 3d
  coordinate generation.
\newblock \emph{PLoS Computational Biology}, 18, 2022.

\bibitem[Engh \& Huber(2012)Engh and Huber]{engh2012structure}
RA~Engh and R~Huber.
\newblock Structure quality and target parameters.
\newblock 2012.

\bibitem[Ferruz et~al.(2022)Ferruz, Schmidt, and
  H{\"o}cker]{ferruz2022protgpt2}
Noelia Ferruz, Steffen Schmidt, and Birte H{\"o}cker.
\newblock Protgpt2 is a deep unsupervised language model for protein design.
\newblock \emph{Nature communications}, 13\penalty0 (1):\penalty0 1--10, 2022.

\bibitem[Fischman \& Ofran(2018)Fischman and Ofran]{fischman2018computational}
Sharon Fischman and Yanay Ofran.
\newblock Computational design of antibodies.
\newblock \emph{Current opinion in structural biology}, 51:\penalty0 156--162,
  2018.

\bibitem[Gao et~al.(2020)Gao, Mahajan, Sulam, and Gray]{gao2020deep}
Wenhao Gao, Sai~Pooja Mahajan, Jeremias Sulam, and Jeffrey~J Gray.
\newblock Deep learning in protein structural modeling and design.
\newblock \emph{Patterns}, 1\penalty0 (9):\penalty0 100142, 2020.

\bibitem[Ho et~al.(2020)Ho, Jain, and Abbeel]{ho2020denoising}
Jonathan Ho, Ajay Jain, and Pieter Abbeel.
\newblock Denoising diffusion probabilistic models.
\newblock \emph{Advances in Neural Information Processing Systems},
  33:\penalty0 6840--6851, 2020.

\bibitem[Hoogeboom et~al.(2022)Hoogeboom, Satorras, Vignac, and
  Welling]{hoogeboom2022equivariant}
Emiel Hoogeboom, Victor~Garcia Satorras, Clement Vignac, and Max Welling.
\newblock Equivariant diffusion for molecule generation in 3d.
\newblock In \emph{International Conference on Machine Learning}, pp.\
  8867--8887. PMLR, 2022.

\bibitem[Hsu et~al.(2022)Hsu, Verkuil, Liu, Lin, Hie, Sercu, Lerer, and
  Rives]{hsu2022learning}
Chloe Hsu, Robert Verkuil, Jason Liu, Zeming Lin, Brian Hie, Tom Sercu, Adam
  Lerer, and Alexander Rives.
\newblock Learning inverse folding from millions of predicted structures.
\newblock \emph{bioRxiv}, 2022.

\bibitem[Huang et~al.(2016)Huang, Boyken, and Baker]{huang2016coming}
Po-Ssu Huang, Scott~E Boyken, and David Baker.
\newblock The coming of age of de novo protein design.
\newblock \emph{Nature}, 537\penalty0 (7620):\penalty0 320--327, 2016.

\bibitem[Ingraham et~al.(2019)Ingraham, Garg, Barzilay, and
  Jaakkola]{ingraham2019generative}
John Ingraham, Vikas Garg, Regina Barzilay, and Tommi Jaakkola.
\newblock Generative models for graph-based protein design.
\newblock \emph{Advances in neural information processing systems}, 32, 2019.

\bibitem[Jia(2008)]{jia2008quaternions}
Yan-Bin Jia.
\newblock Quaternions and rotations.
\newblock \emph{Com S}, 477\penalty0 (577):\penalty0 15, 2008.

\bibitem[Jin et~al.(2021)Jin, Wohlwend, Barzilay, and
  Jaakkola]{jin2021iterative}
Wengong Jin, Jeremy Wohlwend, Regina Barzilay, and Tommi Jaakkola.
\newblock Iterative refinement graph neural network for antibody
  sequence-structure co-design.
\newblock \emph{arXiv preprint arXiv:2110.04624}, 2021.

\bibitem[Jing et~al.(2021)Jing, Eismann, Suriana, Townshend, and
  Dror]{jing2021gvp}
Bowen Jing, Stephan Eismann, Patricia Suriana, Raphael John~Lamarre Townshend,
  and Ron Dror.
\newblock Learning from protein structure with geometric vector perceptrons.
\newblock In \emph{International Conference on Learning Representations}, 2021.

\bibitem[Jumper et~al.(2021)Jumper, Evans, Pritzel, Green, Figurnov,
  Ronneberger, Tunyasuvunakool, Bates, {\v{Z}}{\'\i}dek, Potapenko,
  et~al.]{jumper2021highly}
John Jumper, Richard Evans, Alexander Pritzel, Tim Green, Michael Figurnov,
  Olaf Ronneberger, Kathryn Tunyasuvunakool, Russ Bates, Augustin
  {\v{Z}}{\'\i}dek, Anna Potapenko, et~al.
\newblock Highly accurate protein structure prediction with alphafold.
\newblock \emph{Nature}, 596\penalty0 (7873):\penalty0 583--589, 2021.

\bibitem[Kabsch \& Sander(1983)Kabsch and Sander]{kabsch1983dictionary}
Wolfgang Kabsch and Christian Sander.
\newblock Dictionary of protein secondary structure: pattern recognition of
  hydrogen-bonded and geometrical features.
\newblock \emph{Biopolymers: Original Research on Biomolecules}, 22\penalty0
  (12):\penalty0 2577--2637, 1983.

\bibitem[Kingma \& Welling(2013)Kingma and Welling]{kingma2013auto}
Diederik~P Kingma and Max Welling.
\newblock Auto-encoding variational bayes.
\newblock \emph{arXiv preprint arXiv:1312.6114}, 2013.

\bibitem[Kofinas et~al.(2021)Kofinas, Nagaraja, and Gavves]{kofinas2021roto}
Miltiadis Kofinas, Naveen Nagaraja, and Efstratios Gavves.
\newblock Roto-translated local coordinate frames for interacting dynamical
  systems.
\newblock \emph{Advances in Neural Information Processing Systems},
  34:\penalty0 6417--6429, 2021.

\bibitem[K{\"o}hler et~al.(2020)K{\"o}hler, Klein, and Noe]{kohler20eqflow}
Jonas K{\"o}hler, Leon Klein, and Frank Noe.
\newblock Equivariant flows: Exact likelihood generative learning for symmetric
  densities.
\newblock In \emph{Proceedings of the 37th International Conference on Machine
  Learning}, 2020.

\bibitem[Kong et~al.(2022)Kong, Huang, and Liu]{kong2022conditional}
Xiangzhe Kong, Wenbing Huang, and Yang Liu.
\newblock Conditional antibody design as 3d equivariant graph translation.
\newblock \emph{arXiv preprint arXiv:2208.06073}, 2022.

\bibitem[Leaver-Fay et~al.(2013)Leaver-Fay, O'Meara, Tyka, Jacak, Song,
  Kellogg, Thompson, Davis, Pache, Lyskov, et~al.]{leaver2013scientific}
Andrew Leaver-Fay, Matthew~J O'Meara, Mike Tyka, Ron Jacak, Yifan Song,
  Elizabeth~H Kellogg, James Thompson, Ian~W Davis, Roland~A Pache, Sergey
  Lyskov, et~al.
\newblock Scientific benchmarks for guiding macromolecular energy function
  improvement.
\newblock In \emph{Methods in enzymology}, volume 523, pp.\  109--143.
  Elsevier, 2013.

\bibitem[Luo \& Hu(2021)Luo and Hu]{luo2021diffusion}
Shitong Luo and Wei Hu.
\newblock Diffusion probabilistic models for 3d point cloud generation.
\newblock In \emph{Proceedings of the IEEE/CVF Conference on Computer Vision
  and Pattern Recognition}, pp.\  2837--2845, 2021.

\bibitem[Luo et~al.(2022)Luo, Su, Peng, Wang, Peng, and Ma]{luo2022antigen}
Shitong Luo, Yufeng Su, Xingang Peng, Sheng Wang, Jian Peng, and Jianzhu Ma.
\newblock Antigen-specific antibody design and optimization with
  diffusion-based generative models.
\newblock \emph{bioRxiv}, 2022.

\bibitem[McPartlon \& Xu(2022)McPartlon and Xu]{mcpartlon2022end}
Matt McPartlon and Jinbo Xu.
\newblock An end-to-end deep learning method for rotamer-free protein
  side-chain packing.
\newblock 2022.

\bibitem[Orengo et~al.(1997)Orengo, Michie, Jones, Jones, Swindells, and
  Thornton]{orengo1997cath}
Christine~A Orengo, Alex~D Michie, Susan Jones, David~T Jones, Mark~B
  Swindells, and Janet~M Thornton.
\newblock Cath--a hierarchic classification of protein domain structures.
\newblock \emph{Structure}, 5\penalty0 (8):\penalty0 1093--1109, 1997.

\bibitem[Rives et~al.(2021)Rives, Meier, Sercu, Goyal, Lin, Liu, Guo, Ott,
  Zitnick, Ma, et~al.]{rives2021biological}
Alexander Rives, Joshua Meier, Tom Sercu, Siddharth Goyal, Zeming Lin, Jason
  Liu, Demi Guo, Myle Ott, C~Lawrence Zitnick, Jerry Ma, et~al.
\newblock Biological structure and function emerge from scaling unsupervised
  learning to 250 million protein sequences.
\newblock \emph{Proceedings of the National Academy of Sciences}, 118\penalty0
  (15):\penalty0 e2016239118, 2021.

\bibitem[Ruffolo et~al.(2022)Ruffolo, Sulam, and Gray]{ruffolo2022antibody}
Jeffrey~A Ruffolo, Jeremias Sulam, and Jeffrey~J Gray.
\newblock Antibody structure prediction using interpretable deep learning.
\newblock \emph{Patterns}, 3\penalty0 (2):\penalty0 100406, 2022.

\bibitem[Saka et~al.(2021)Saka, Kakuzaki, Metsugi, Kashiwagi, Yoshida, Wada,
  Tsunoda, and Teramoto]{saka2021antibody}
Koichiro Saka, Taro Kakuzaki, Shoichi Metsugi, Daiki Kashiwagi, Kenji Yoshida,
  Manabu Wada, Hiroyuki Tsunoda, and Reiji Teramoto.
\newblock Antibody design using lstm based deep generative model from phage
  display library for affinity maturation.
\newblock \emph{Scientific reports}, 11\penalty0 (1):\penalty0 1--13, 2021.

\bibitem[Shen et~al.(2018)Shen, Fallas, Lynch, Sheffler, Parry, Jannetty,
  Decarreau, Wagenbach, Vicente, Chen, et~al.]{shen2018novo}
Hao Shen, Jorge~A Fallas, Eric Lynch, William Sheffler, Bradley Parry, Nicholas
  Jannetty, Justin Decarreau, Michael Wagenbach, Juan~Jesus Vicente, Jiajun
  Chen, et~al.
\newblock De novo design of self-assembling helical protein filaments.
\newblock \emph{Science}, 362\penalty0 (6415):\penalty0 705--709, 2018.

\bibitem[Shi et~al.(2021)Shi, Luo, Xu, and Tang]{shi2021learning}
Chence Shi, Shitong Luo, Minkai Xu, and Jian Tang.
\newblock Learning gradient fields for molecular conformation generation.
\newblock In \emph{International Conference on Machine Learning}, pp.\
  9558--9568. PMLR, 2021.

\bibitem[Shin et~al.(2021)Shin, Riesselman, Kollasch, McMahon, Simon, Sander,
  Manglik, Kruse, and Marks]{shin2021protein}
Jung-Eun Shin, Adam~J Riesselman, Aaron~W Kollasch, Conor McMahon, Elana Simon,
  Chris Sander, Aashish Manglik, Andrew~C Kruse, and Debora~S Marks.
\newblock Protein design and variant prediction using autoregressive generative
  models.
\newblock \emph{Nature communications}, 12\penalty0 (1):\penalty0 1--11, 2021.

\bibitem[Song \& Ermon(2019)Song and Ermon]{song2019generative}
Yang Song and Stefano Ermon.
\newblock Generative modeling by estimating gradients of the data distribution.
\newblock \emph{Advances in Neural Information Processing Systems}, 32, 2019.

\bibitem[Steinegger \& S{\"o}ding(2017)Steinegger and
  S{\"o}ding]{steinegger2017mmseqs2}
Martin Steinegger and Johannes S{\"o}ding.
\newblock Mmseqs2 enables sensitive protein sequence searching for the analysis
  of massive data sets.
\newblock \emph{Nature biotechnology}, 35\penalty0 (11):\penalty0 1026--1028,
  2017.

\bibitem[Tischer et~al.(2020)Tischer, Lisanza, Wang, Dong, Anishchenko, Milles,
  Ovchinnikov, and Baker]{tischer2020design}
Doug Tischer, Sidney Lisanza, Jue Wang, Runze Dong, Ivan Anishchenko, Lukas~F
  Milles, Sergey Ovchinnikov, and David Baker.
\newblock Design of proteins presenting discontinuous functional sites using
  deep learning.
\newblock \emph{Biorxiv}, 2020.

\bibitem[Trippe et~al.(2022)Trippe, Yim, Tischer, Broderick, Baker, Barzilay,
  and Jaakkola]{trippe2022diffusion}
Brian~L Trippe, Jason Yim, Doug Tischer, Tamara Broderick, David Baker, Regina
  Barzilay, and Tommi Jaakkola.
\newblock Diffusion probabilistic modeling of protein backbones in 3d for the
  motif-scaffolding problem.
\newblock \emph{arXiv preprint arXiv:2206.04119}, 2022.

\bibitem[Tubiana et~al.(2022)Tubiana, Schneidman-Duhovny, and
  Wolfson]{tubiana2022scannet}
J{\'e}r{\^o}me Tubiana, Dina Schneidman-Duhovny, and Haim~J Wolfson.
\newblock Scannet: An interpretable geometric deep learning model for
  structure-based protein binding site prediction.
\newblock \emph{Nature Methods}, pp.\  1--10, 2022.

\bibitem[van Kempen et~al.(2022)van Kempen, Kim, Tumescheit, Mirdita,
  S{\"o}ding, and Steinegger]{van2022foldseek}
Michel van Kempen, Stephanie Kim, Charlotte Tumescheit, Milot Mirdita, Johannes
  S{\"o}ding, and Martin Steinegger.
\newblock Foldseek: fast and accurate protein structure search.
\newblock \emph{bioRxiv}, 2022.

\bibitem[Vaswani et~al.(2017)Vaswani, Shazeer, Parmar, Uszkoreit, Jones, Gomez,
  Kaiser, and Polosukhin]{vaswani2017attention}
Ashish Vaswani, Noam Shazeer, Niki Parmar, Jakob Uszkoreit, Llion Jones,
  Aidan~N Gomez, {\L}ukasz Kaiser, and Illia Polosukhin.
\newblock Attention is all you need.
\newblock \emph{Advances in neural information processing systems}, 30, 2017.

\bibitem[Wang et~al.(2021)Wang, Lisanza, Juergens, Tischer, Anishchenko, Baek,
  Watson, Chun, Milles, Dauparas, et~al.]{wang2021deep}
Jue Wang, Sidney Lisanza, David Juergens, Doug Tischer, Ivan Anishchenko,
  Minkyung Baek, Joseph~L Watson, Jung~Ho Chun, Lukas~F Milles, Justas
  Dauparas, et~al.
\newblock Deep learning methods for designing proteins scaffolding functional
  sites.
\newblock \emph{bioRxiv}, 2021.

\bibitem[Xu et~al.(2022)Xu, Yu, Song, Shi, Ermon, and Tang]{xu2022geodiff}
Minkai Xu, Lantao Yu, Yang Song, Chence Shi, Stefano Ermon, and Jian Tang.
\newblock Geodiff: A geometric diffusion model for molecular conformation
  generation.
\newblock \emph{arXiv preprint arXiv:2203.02923}, 2022.

\bibitem[Yang et~al.(2020)Yang, Anishchenko, Park, Peng, Ovchinnikov, and
  Baker]{yang2020improved}
Jianyi Yang, Ivan Anishchenko, Hahnbeom Park, Zhenling Peng, Sergey
  Ovchinnikov, and David Baker.
\newblock Improved protein structure prediction using predicted interresidue
  orientations.
\newblock \emph{Proceedings of the National Academy of Sciences}, 117\penalty0
  (3):\penalty0 1496--1503, 2020.

\bibitem[Zhu et~al.(2022)Zhu, Xia, Liu, Wu, Xie, Wang, Wang, Zhou, Qin, Li,
  et~al.]{zhu2022direct}
Jinhua Zhu, Yingce Xia, Chang Liu, Lijun Wu, Shufang Xie, Tong Wang, Yusong
  Wang, Wengang Zhou, Tao Qin, Houqiang Li, et~al.
\newblock Direct molecular conformation generation.
\newblock \emph{arXiv preprint arXiv:2202.01356}, 2022.

\end{thebibliography}
\bibliographystyle{iclr2023_conference}

\clearpage
\appendix


\section{Equivariance Property of Sequence and Structure Translation}
We first quickly recap the process of sequence and structure translation in each translation layer.
At $(t+1)^{\mathrm{th}}$ layer, the decoder takes protein $\gP^t = \{(\vs_i^t, \vx_i^t, \mO_i^t)\}_{i=1}^N$ and context features $\{\vm_i^t\}, \{\vz_{ij}^t\}$ as the input.
It encodes sequence-structure interplay and integrates all interactions into updated context features using $\operatorname{SeqIPA}$ adapted from Invariant Point Attention (IPA)~\citep{jumper2021highly} in a way that its roto-translation invariant property is kept.
The updates of C$_\alpha$ positions, frame orientations, and type distributions are then predicted based on updated context features.
The whole process can be summarized as follows:
\begin{align}
\vs_i^{t+0.5} 
&= \operatorname{MLP_e}(\vs_i^t), \\
\label{eq:suppl_seqipa}
\vm_i^{t+1}, \vz_{ij}^{t+1}
&= \operatorname{SeqIPA}(\{\vm_i^{t}\}, \{\vz_{ij}^{t}\}, \{\vs_i^{t+0.5}\}, \{\vx_i^t\}, \{\mO_i^t\}),\\
\label{eq:suppl_ca_update}
\hat{\vx}_i^t
&= \operatorname{MLP_x}(\vm_i^{t+1}, \vm_i^{0}),
\quad \vx_i^{t+1} = \vx_i^{t} + \Delta{\vx_i^t} = \vx_i^{t} + \mO_i^t \hat{\vx}_i^t, \\
\label{eq:suppl_frame_update}
\hat{\mO}_i^t
&= \operatorname{convert}\big(\operatorname{MLP_o}(\vm_i^{t+1}, \vm_i^{0})\big),
\quad \mO_i^{t+1} = \mO_i^t \hat{\mO}_i^t, \\
\label{eq:suppl_type_update}
\vs_i^{t+1}
&= \operatorname{softmax} \big(\lambda \cdot \operatorname{MLP_s}(\vm_i^{t+1}, \vm_i^{0}, \vs_i^{t+0.5})\big ).
\end{align}
To derive the equivariance property of each translation step, we use three functions $\gX, \gO, \gS$ to denote the network that predicts the C$_\alpha$ position translation, orientation translation, and sequence translation described above, respectively.
Formally, we have:
\begin{align}
\Delta\vx_i^{t} &= \gX(\gP^t), \\
\mO_i^{t+1} &= \gO(\gP^t), \\
\vs_i^{t+1} &= \gS(\gP^t).
\end{align}
Note that $\gX, \gO, \gS$ also take $\{\vm_i^{t}\}$ and $\{\vz_{ij}^{t}\}$ as input and we omit these context features for simplicty, as they remain invariant to global rigid transformations.
$\gX, \gO, \gS$ are not separate networks, and they share the same input and the same $\operatorname{SeqIPA}$, but are equipped with different MLPs.
With the above definitions, we can derive the following proposition:
\begin{proposition}[Roto-Translation Equivariance]
\label{prop:equivariance}
Let $\gT_{\mR, \vr}$ denote any SE(3) transformation (rigid transformation) operating on the protein object $\gP^t=\{(\vs_i^t, \vx_i^t, \mO_i^t)\}_{i=1}^N$, with a rotation matrix $\mR \in \mathrm{SO(3)}$ and a translation vector $\vr \in \sR^3$. The
function $\gX, \gO, \gS$ satisfy the following equivariance properties:
\begin{align}
    \label{eq:suppl_equi_x}
    \gX \circ \gT_{\mR, \vr} (\gP^t) &= \mR \gX (\gP^t),\\
    \label{eq:suppl_equi_o}
    \gO \circ \gT_{\mR, \vr} (\gP^t) &= \mR \gO (\gP^t),\\
    \label{eq:suppl_equi_s}
    \gS \circ \gT_{\mR, \vr} (\gP^t) &= \gS (\gP^t),
\end{align}
where $\gT_{\mR, \vr}(\gP^t) = \{(\vs_i^t, \vx_i^t + \vr, \mR\mO_i^t )\}_{i=1}^N$.
\end{proposition}
Intuitively, the proposition states that in each translation step, the updates of C$_\alpha$ positions and frame orientations are equivariant with respect to input protein structures, and the updates of type distributions are invariant.

\begin{proof}
We first prove that Eq.~\ref{eq:suppl_equi_x} holds.
Notice that $\operatorname{SeqIPA}$ is aware of the orientations of the input structure, and the updated context features are invariant (Eq.~\ref{eq:suppl_seqipa}).
Therefore, the predicted deviation of C$_\alpha$ positions, i.e., $\hat{\vx}_i^t$, is invariant (Eq.~\ref{eq:suppl_ca_update}).
Then, we have:
\begin{align}
    \gX \circ \gT_{\mR, \vr} (\gP^t) 
    &= \mR \mO_i^t \hat{\vx}_i^t = \mR \gX (\gP^t).
\end{align}
The Eq.~\ref{eq:suppl_equi_o} and Eq.~\ref{eq:suppl_equi_s} can be proved in a similar way.
\end{proof}
\clearpage
\section{Model Details}
\label{sec:suppl_model_details}
\subsection{Pseudo Code}
\label{subsec:suppl_pseudo_code}
The pseudo code of \method is provided in Algorithm~\ref{alg:method}.
The proposed \method consists of a trigonometry-aware encoder (Algorithm~\ref{alg:method}, line~\ref{alg:encoder_start}-\ref{alg:encoder_end}) that reasons geometrical constraints and
interactions from context features, and a roto-translation equivariant decoder (Algorithm~\ref{alg:method}, line~\ref{alg:decoder_start}-\ref{alg:decoder_end}) that translates protein sequence and structure interdependently. 
Starting from the intial single features $\{\vm_i\} \in \sR^{N \times {c_m}}$ and pair features $\{\vz_{ij}\} \in \sR^ {N \times N \times {c_z}}$, the whole model iteratively translates both protein sequence and structure into the desired state
from random initialization (Algorithm~\ref{alg:method}, line~\ref{alg:initialize_p}). We note that the whole process does not require MCMC sampling, and runs much faster than autoregressive models and diffusion-based models.

The trigonometry-aware encoder is composed of a stack of $L$ encoding layers.
Each layer takes single features $\{\vm_i^l\}$ and pair features $\{\vz_{ij}^l\}$ from the last layer as its input, and 
updates these features with novel attention mechanisms.
After $L$ rounds of feature propagation, the updated single features and pair features serve as inputs to the decoder
for joint protein sequence-structure design (Algorithm~\ref{alg:method}, line~\ref{alg:initialize_c}).

The roto-translation equivariant decoder iteratively refines the concrete 3D atom coordinates of the protein from random initialization, based on context features calculated by the encoder. 
The interplay of residue types, residue structures and context features during the decoding process is captured by a novel orientation-aware attention mechanism ($\operatorname{SeqIPA}$, Appendix~\ref{subsec:suppl_seqipa}).
The equivariance property of the structure translation process is guaranteed by mapping invariant predictions in local frames to global frames with change of basis (Algorithm~\ref{alg:method}, line~\ref{alg:translation_start}-\ref{alg:translation_end}).
It is worth mentioning that \method updates sequence and structure of all residues in an one-shot manner, leading to a much more efficient inference process (Algorithm~\ref{alg:method}, line~\ref{alg:seq_update}).

\begin{algorithm}[H]

	\caption{\method}
	\label{alg:method}
    \renewcommand\algorithmiccomment[1]{\hfill $\triangleright$ {#1}}
    \renewcommand{\algorithmicrequire}{\textbf{Require:}}
    \renewcommand{\algorithmicensure}{\textbf{Return:}}
\begin{algorithmic}[1]

    \REQUIRE{Initial single features $\{\vm_i\} \in \sR^{N \times {c_m}}$ and pair features $\{\vz_{ij}\} \in \sR^ {N \times N \times {c_z}}$.}
    \STATE $\vm_i^0, \vz_{ij}^0 \gets \operatorname{Linear}(\vm_i), \operatorname{Linear}(\vz_{ij})$
    \label{alg:initial_context}
    \COMMENT{$\vm_i^0 \in \sR^{c}, \vz_{ij}^0 \in \sR^{c}$}
    \FOR{$l \gets 0$ to $L-1$}
    \label{alg:encoder_start}
        \STATE{$\vm_i^{l+1} \gets
        \operatorname{MHA}(\{\vm_i^l\}, \{\vz_{ij}^l\})$}
        \label{alg:one}
        \COMMENT{Eq.~\ref{eq:res_update}}
        \STATE{$\vz_{ij}^{l+0.5} \gets \vz_{ij}^{l} + \operatorname{Linear}(\vm_i^{l+1} \otimes \vm_j^{l+1})$}
        \COMMENT{Eq.~\ref{eq:res2pair}}
        \STATE{$\vz_{ij}^{l+0.75} \gets \vz_{ij}^{l+0.5} + \operatorname{TriangleUpdate_1}(\{\vz_{ij}^{l+0.5})\}$}
        \COMMENT{Eq.~\ref{eq:triangle_mul}}
        \STATE{$\vz_{ij}^{l+1} \gets \vz_{ij}^{l+0.75} + \operatorname{TriangleUpdate_2}(\{\vz_{ij}^{l+0.75})\}$}
        \COMMENT{Eq.~\ref{eq:triangle_att}}
    \ENDFOR
    \label{alg:encoder_end}
    \STATE $\vm_i^0, \vz_{ij}^0 \gets \vm_i^L, \vz_{ij}^L$
    \label{alg:initialize_c}
    \COMMENT{Initialize context features for decoder}
    \STATE $\gP^0 \gets \{(\vs_i^0, \vx_i^0, \mO_i^0) \}_{i=1}^N
                  \gets \{(\frac{1}{20} \cdot \bm{1}, (0, 0, 0), \mI_{\bm{3}}) \}_{i=1}^N
           $
    \label{alg:initialize_p}
    \COMMENT{Initialize protein $\gP^0$}
    \FOR{$t \gets 0$ to $T-1$}
    \label{alg:decoder_start}
    \STATE $\vs_i^{t+0.5} \gets \operatorname{MLP_e}(\vs_i^t)$
    \COMMENT{$\vs_i^{t+0.5} \in \sR^c$}
    \STATE $\vm_i^{t+1}, \vz_{ij}^{t+1} \gets \operatorname{SeqIPA}(\{\vm_i^{t}\}, \{\vz_{ij}^{t}\}, \{\vs_i^{t+0.5}\}, \{\vx_i^t\}, \{\mO_i^t\})$
    \COMMENT{Eq.~\ref{eq:seqipa} and Section~\ref{subsec:suppl_seqipa}}
    \STATE $\hat{\vx}_i^t \gets \operatorname{MLP_x}(\vm_i^{t+1}, \vm_i^{0}) $
    \label{alg:translation_start}
    \COMMENT{Deviation of C$_\alpha$ positions in local frame}
    \STATE $\vx_i^{t+1} \gets \vx_i^{t} + \mO_i^t \hat{\vx}_i^t $
    \COMMENT{Deviation of C$_\alpha$ positions in global frame}
    \STATE $\hat{\mO}_i^t \gets \operatorname{convert}\big(\operatorname{MLP_o}(\vm_i^{t+1}, \vm_i^{0})\big)$
    \COMMENT{Convert a quaternion to a rotation matrix}
    \STATE $\mO_i^{t+1} \gets \mO_i^t \hat{\mO}_i^t$
    \label{alg:translation_end}
    \STATE $\vs_i^{t+1} \gets \operatorname{softmax} \big(\lambda \cdot \operatorname{MLP_s}(\vm_i^{t+1}, \vm_i^{0}, \vs_i^{t+0.5})\big)$
    \label{alg:seq_update}
    \COMMENT{Eq.~\ref{eq:seq_update}}
    \STATE $\gP^{t+1} \gets \{(\vs_i^{t+1}, \vx_i^{t+1}, \mO_i^{t+1}) \}_{i=1}^N$
    \ENDFOR
    \label{alg:decoder_end}
    \ENSURE{The trajectory of the protein translation $\{\gP^t\}_{t=1}^T$.}

\end{algorithmic}
\end{algorithm}

\clearpage
\subsection{Parameterization of SeqIPA}
\label{subsec:suppl_seqipa}
$\operatorname{SeqIPA}$ is adapted from the Invariant Point Attention (IPA)~\citep{jumper2021highly}, which takes residue types as the additional input to capture the interactions between current decoded sequences, structures, and the context features. We ensure that the additional input does not affect the invariance property of the IPA to make full use of its capacity.
Specifically, we propose the following two strategies to parameterize the $\operatorname{SeqIPA}$.

\textbf{SeqIPA-Addition.}
Given that $\{\vs_i^{t+0.5}\}$ share the same dimensionality with $\{\vm_i^t\}$, a very simple strategy is to just add embeddings of residue types onto single representations.
Following the original implementation of IPA, we leave the pair features unchanged in this approach.
\begin{align}
\label{eq:suppl_seqipa_addition}
\vm_i^{t+1}, \vz_{ij}^{t+1}
&= \operatorname{SeqIPA}(\{\vm_i^{t}\}, \{\vz_{ij}^{t}\}, \{\vs_i^{t+0.5}\}, \{\vx_i^t\}, \{\mO_i^t\}) \\
\label{eq:suppl_seqipa_addition2}
&= \operatorname{IPA}(\{\vm_i^{t} + \vs_i^{t+0.5}\}, \{\vz_{ij}^{t}\}, \{\vx_i^t\}, \{\mO_i^t\}).
\end{align}
The above equations say that for SeqIPA-Addition, we just add the sequence embeddings $\{\vs_i^{t+0.5}\}$ onto the single representation $\{\vm_i^t\}$ (Eq.\ref{eq:suppl_seqipa_addition2}), and feed four inputs to vanilla IPA~\citep{jumper2021highly}.

\textbf{SeqIPA-Attention.}
Another more complicated strategy is to construct a new set of single representations and pair representations based on the embeddings of the current residue types.
Then, we adopt a lightweight encoder similar to the encoder introduced in Section~\ref{subsec:encoder} to update $\vm_i^t$ and $\vz_{ij}^t$, which are then fed into the vanilla IPA module. We summarize the computation flow as follows:
\begin{align}
\label{eq:suppl_seqipa_attention}
    \bar{\vm}_i, \bar{\vz}_{ij} &= \operatorname{Linear}(\vs_i^{t+0.5}), \operatorname{Linear}(\vs_i^{t+0.5} + \vs_j^{t+0.5}) \\
\label{eq:suppl_seqipa_attention2}
    \bar{\vm}_i, \bar{\vz}_{ij} &= \operatorname{Encoder}(\{\bar{\vm}_i\}, \{\bar{\vz}_{ij}\}), \\
\label{eq:suppl_seqipa_attention3}
    \vm_i^{t+0.5}, \vz_{ij}^{t+0.5} &= \vm_i^t + \bar{\vm}_i, \vz_{ij}^t + \bar{\vz}_{ij}, \\
    \vm_i^{t+1}, \vz_{ij}^{t+1}
&= \operatorname{SeqIPA}(\{\vm_i^{t}\}, \{\vz_{ij}^{t}\}, \{\vs_i^{t+0.5}\}, \{\vx_i^t\}, \{\mO_i^t\}) \\
\label{eq:suppl_seqipa_attention4}
&= \operatorname{IPA}(\{\vm_i^{t+0.5}\}, \{\vz_{ij}^{t+0.5}\}, \{\vx_i^t\}, \{\mO_i^t\}).
\end{align}
The above equations say that for SeqIPA-Attention, we leverage another lightweight encoder (Eq.\ref{eq:suppl_seqipa_attention2}) similar to the encoder introduced in Section~\ref{subsec:encoder} to first update $\vm_i^t$ and $\vz_{ij}^t$ (Eq.\ref{eq:suppl_seqipa_attention3}), and then feed updated inputs to vanilla IPA (Eq.\ref{eq:suppl_seqipa_attention4}).

In practice, we find both strategies work well and their performance is on par with each other.
To make the whole model lightweight, we adopt the first strategy across all the experiments in this work.
We emphasize that the parameterization of the $\operatorname{SeqIPA}$ is quite flexible, as long as it can model interactions between sequences, structures, and context features, and is invariant to the global transformation of input structures. For the concrete computation flows of the IPA module, we refer readers to Algorithm 22 described in the supplementary material of~\cite{jumper2021highly}.

\subsection{Hyper-parameters and Implementation Details}
\label{subsec:hyper_parameters}

\method is implemented in Pytorch.
The trigonometry-aware context encoder is implemented with $L=8$ layers, and the sequence-structure decoder is implemented with $T=8$ layers.
The hidden dimension is set as 128 for pair features and 256 for single features across all modules.

For training, we use a learning rate of 0.001 with 2000 linear warmup iterations.
We empirically find that proper learning rate warmup schedule can lead to faster convergence rate and higher performance.
The model is optimized with Adam optimizer on four Tesla V100 GPU cards with distributed data parallel.
The estimated time it takes to get a converged model is 24 hours.

For inference, the temperature of the sequence distribution, i.e., $\lambda$, controls the sharpness of the distribution.
The larger $\lambda$ will lead to higher (better) AAR and higher (worse) PPL, and vice versa. Since it acts oppositely on AAR and PPL, we simply set it as 1 across the experiments.

All codes, datasets, and experimental environments will be released upon the acceptance of this work.




\subsection{Full atom positions reconstruction}
The bond lengths and bond angles between backbone atoms are relatively conserved.
The C$_\alpha$ is connected with N, C, and C$_\beta$ (except for Glycine which has a single hydrogen atom as its side chain) atoms. The C$_\alpha$ forms a canonical orientation frame with respect to N, C, and C$_\beta$.
Once we know the positions of the C$_\alpha$, and the orientation of the frame, the full backbone atom positions can be derived according to their averaged relative positions with respect to the C$_\alpha$ recorded in literature.
The positions of all sidechain atoms of the 20 different amino acids can also be compactly specified by four torsion angles ($\chi_1, \chi_2, \chi_3, \chi_4$)~\citep{jumper2021highly,mcpartlon2022end}, which also follow amino-acid specific distributions recorded in literature.
All these recorded statistics can be found in
\url{https://git.scicore.unibas.ch/schwede/openstructure/-/raw/7102c63615b64735c4941278d92b554ec94415f8/modules/mol/alg/src/stereo_chemical_props.txt}.
\clearpage
\section{Experimental Details}

\begin{figure*}[t]
	\centering
	\vspace{-20pt}
    \includegraphics[width=1.0\linewidth]{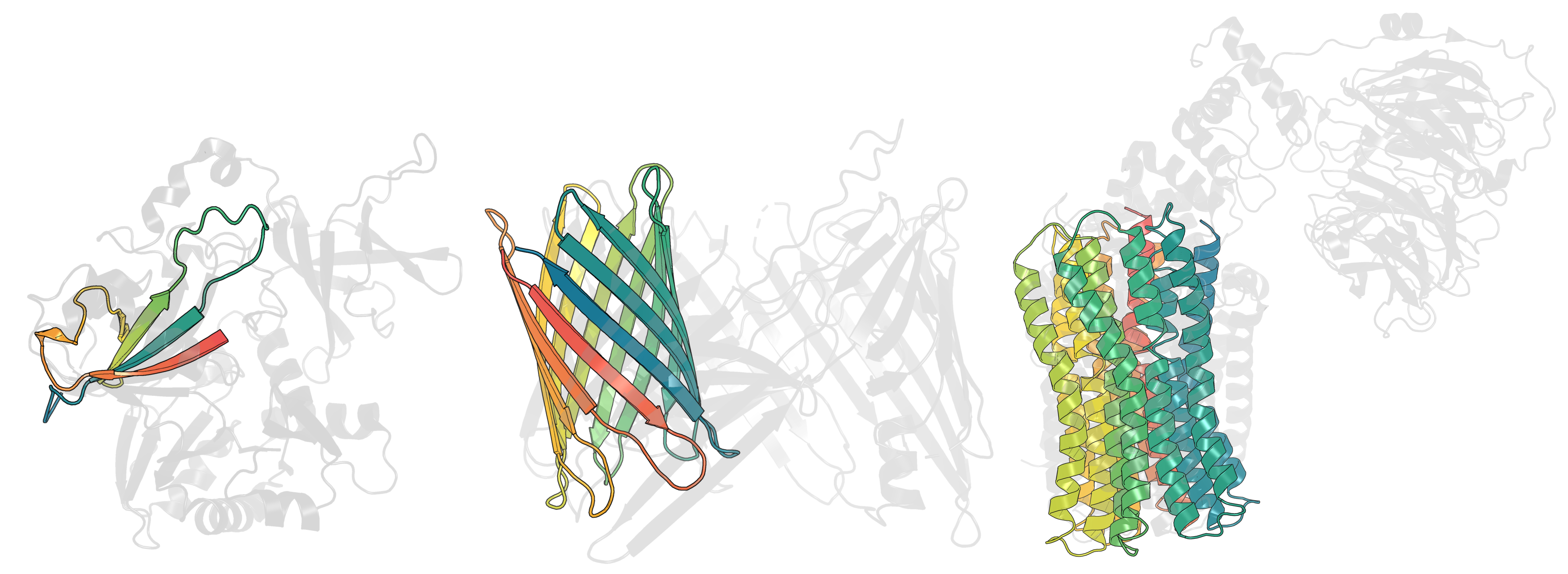}
    \vspace{-10pt}
    \caption{
    Superimposition of three generated proteins and their most similar proteins found in the PDB by FoldSeek.
    left: a novel protein with loop extended.
    middle: a novel $\beta-$barrel.
    right: a novel helical complex.
    }
    \label{fig:suppl_case_study}
    \vspace{-10pt}
\end{figure*}

\subsection{Case Study}
\label{subsec:case_study}
We conduct three case studies to evaluate \method's capability to perform \textit{de novo} protein design, including extending the loop of existing proteins, designing novel $\beta-$barrels, and designing novel helical complexes.
Specifically, we manually curate a set of secondary structure annotations and contact features, and ask the model trained in the second task (Section~\ref{sec:prot_design}) to generate novel proteins based on these context features.
We elaborate the way we design context features for each setting.

\textbf{Extending the Loop.}
In this setting, we start by calculating the secondary structure annotations and the contact matrix of an existing protein.
We then insert $n$ consecutive ``C'' (``C'' is the secondary structure annotation for the loop) letters into the original secondary structure annotations at the position where we want to extend the loop.
Similarly, we insert $n$ consecutive rows and columns filled with zero into the original contact matrix.
For a new-inserted residue indexed by $i$, we let it to be in contact with $i-2, i-1, i, i+1, i+2$.

\textbf{Designing Novel Beta Barrels.}
In this setting, we grab a simple pattern of $\beta-$barrel proteins and then repeat this pattern multiple times to construct the contact features.
The secondary structure annotations are also calculated by repeating the annotations of the pattern multiple times.

\textbf{Designing Novel Helical Complexes.}
Similar to the second case, in this setting, we also take a simple pattern of helical complexes and construct the contact features by repeating it multiple times.
The secondary structure annotations are all set to be ``H''.

In Figure~\ref{fig:suppl_case_study}, we show the superimposition of three novel proteins designed by \method against the most similar proteins in the PDB, one for each setting, which confirms the novelty of the designed proteins.

\clearpage
\subsection{Ablation Study}
To gain more insights into the effectiveness of each module in \method, we conduct additional ablation studies following the setting of the antibody CDR co-design in Section~\ref{sec:ab_design}.

\textbf{Effectiveness of Cross-conditioning on Sequence and Structure.}
In this ablation study (denoted as \textbf{No Sequence Interaction}), we replace the $\operatorname{SeqIPA}$ in Eq.\ref{eq:seqipa} with the vanilla $\operatorname{IPA}$~\citep{jumper2021highly}, and directly predict the distribution of amino acid types for all residues at the last iteration of the decoding process using single features, i.e., $\{\vm_i^T\}$.
The results shown in Table~\ref{tab:ablation} indicate that when the model fails to cross-condition on both sequence and structure during the decoding, there is a significant performance drop in all three metrics, especially for PPL and AAR.
This confirms the necessity to cross-condition on sequence and structure during the decoding, and the effectiveness of the proposed $\operatorname{SeqIPA}$.

\textbf{Effectiveness of Iterative Translations.}
In this ablation study (denoted as \textbf{Single Iteration Translation}), we replace the $T-$layer decoder with a single-layer decoder for protein translation. We note that the vanilla decoder of \method is composed of $T = 8$ consecutive translation layers with tied weights, and the number of trainable parameters of these two models are the same due to the weight tying.
As indicated by Table~\ref{tab:ablation}, the single-layer decoder is outperformed by the vanilla decoder by a large margin on all metrics. Since the numbers of trainable parameters of these two models are the same, this is a fair comparison. The results justify the advantages of the iterative translation framework for protein sequence and structure co-design.

\textbf{Effectiveness of Context Feature Update.}
In this ablation study (denoted as \textbf{No SeqIPA}), we freeze all context feature updates in the decoder and remove $\operatorname{SeqIPA}$ (Eq.\ref{eq:seqipa}), and use the outputs of the encoder as the context features during the whole decoding process.
As shown by Table~\ref{tab:ablation}, the performance degrades dramatically, which demonstrates that the context feature update plays a key role in \method.

\vspace{-5pt}
\begin{table}[t]
    \centering
    \vspace{-5pt}
    \caption{
        Ablation study on the antibody CDR co-design task. ($\uparrow$): the higher the better. ($\downarrow$): the lower the better.
    }
    \label{tab:ablation}
    \vspace{2pt}  

    \begin{adjustbox}{width=1.0\linewidth}
        \begin{tabular}{l | ccc | ccc | ccc}
\toprule
& \multicolumn{3}{c|}{PPL ($\downarrow$)} 
& \multicolumn{3}{c|}{RMSD (\AA, $\downarrow$)} 
& \multicolumn{3}{c}{AAR (\%, $\uparrow$)} \\

CDR 
& H1 & H2 & H3
& H1 & H2 & H3
& H1 & H2 & H3 \\
\midrule
\textbf{\method}
& \textbf{4.43}  & \textbf{5.94}  & \textbf{10.88}
& \textbf{1.24}  & \textbf{1.11}  & \textbf{3.19}
& \textbf{70.22}  & \textbf{63.53}  & \textbf{39.27} \\
\midrule
No Sequence Interaction
& 5.19 & 7.77 & 14.78
& 1.48 & 1.32 & 3.24
& 68.16 & 59.14 & 32.42 \\
Single Iteration Translation
& 6.03 & 17.98 & 26.11
& 2.17 & 2.41 & 5.09
& 67.49 & 55.82 & 29.42 \\
No SeqIPA
& 13.96 & 14.57 & 17.87 
& 18.22 & 18.92 & 11.37 
& 9.88 & 10.14 & 7.42 \\
\bottomrule
\end{tabular}

    \end{adjustbox}
    \vspace{-5pt}
\end{table}

\clearpage
\section{Discussion}

\subsection{Context features}

In this work, we use the concept of context features (single features $\{\vm_i\}$ and pair features $\{\vz_{ij}\}$) as a formulation to unify the inputs of different protein design tasks. 
We note that $\{\vm_i\}$ and $\{\vz_{ij}\}$ vary from task to task, and for most well-defined tasks, they are easy to get.
For example, in antibody CDR design tasks~\citep{jin2021iterative, luo2022antigen, kong2022conditional}, they can be derived from antibody frameworks and structures of binding antigens.
In the general protein design task proposed by~\cite{anand2022protein}, they can be derived from second structure annotations and contact maps provided by biologists. In scaffolding tasks, they can be derived from an starting motif~\citep{trippe2022diffusion}.
The more context features (or constraints) the researchers specify, the more control they can have over the designed proteins.
We refer readers to~\cite{dou2018novo, shen2018novo} for two cases of protein design in real world scenarios.

\subsection{Context-free protein design}

Researchers may be interested at generating novel proteins without relying on any context features, a.k.a. context-free protein design. We note that \method is a very general framework and can handle this situation with minor modifications.
Specifically, we can adopt the Variational Autoencoder (VAE) framework~\citep{kingma2013auto} and approximate the data distribution by learning to map proteins to latent vectors and reconstruct proteins from latents. In this scenario, the context features become the proteins sampled from target data distributions.
Since this is out of the scope of this work, we leave it as our future work.

\end{document}